\documentclass[a4paper,12pt]{article}
\pdfoutput=1

\usepackage[titletoc,title]{appendix}

\usepackage{test}

\usepackage{hyperref}
\hypersetup{colorlinks,citecolor=blue,urlcolor=magenta}
\usepackage{doi}

\usepackage{mathrsfs}
\usepackage{bbm}

\usepackage{booktabs,caption}

\usepackage[style=ext-authoryear-comp,
sorting=nyt,
dashed=false, 
maxcitenames=2, 
maxbibnames=99, 
uniquelist=false,
uniquename=false,
useprefix=true,
giveninits=true, 
natbib, 
date=year 
]{biblatex}

\AtBeginRefsection{\GenRefcontextData{sorting=ynt}}
\AtEveryCite{\localrefcontext[sorting=ynt]}

\DeclareFieldFormat{pages}{#1} 
\renewbibmacro{in:}{\ifentrytype{article}{}{\printtext{\bibstring{in}\intitlepunct}}} 

\DeclareFieldFormat[article,inbook,incollection,inproceedings,patent,thesis,unpublished]{titlecase:title}{\MakeSentenceCase*{#1}} 

\usepackage[inline,shortlabels]{enumitem}
\setlist[enumerate,1]{
    label=(\roman*),
    itemsep=2pt,
    topsep=4pt,
    parsep=0pt,
    partopsep=0pt
}

\usepackage[margin = 1.25in]{geometry}
\usepackage{setspace}
\title{Wealth Preferences and the Upper Tail of Consumption\thanks{Ma acknowledges financial support from the National Natural Science Foundation of China (Grant ID: 72573111). Toda acknowledges financial support from Japan Center for Economic Research; Japan Securities Scholarship Foundation; and the Joint Usage/Research Center, Institute of Economic Research, Hitotsubashi University (Grant ID: IERPK2515).}}
\author{Qingyin Ma\thanks{International School of Economics and Management, Capital University of Economics and Business. Email: \href{mailto:qingyin.ma@cueb.edu.cn}{qingyin.ma@cueb.edu.cn}.} \and Alexis Akira Toda\thanks{Department of Economics, Emory University and Research Institute for Economics and Business Administration, Kobe University. Email: \href{mailto:alexis.akira.toda@emory.edu}{alexis.akira.toda@emory.edu}.}}

\numberwithin{lem}{section}
\numberwithin{prop}{section}

\renewcommand{\AA}{\mathsf A}
\renewcommand{\SS}{\mathsf S}
\newcommand{\XX}{\mathsf X}
\newcommand{\ZZ}{\mathsf Z}
\renewcommand{\Pr}{\operatorname{P}} 
\newcommand{\cC}{\mathscr C}
\newcommand{\fF}{\mathscr F}
\newcommand{\hH}{\mathscr H}
\newcommand{\iidsim}{\stackrel {\textrm{ {\sc iid }}} {\sim} }

\newcommand{\psihat}{0.1447}

\begin{document}
\maketitle

\begin{abstract}

We develop a theory of optimal saving when wealth enters utility. Relative curvatures of consumption and wealth utility, $\gamma$ and $\delta$, govern the upper-tail behavior. When $\delta<\gamma$, wealth preferences generate a vanishing asymptotic marginal propensity to consume and power-law consumption, $c(w)\sim w^{\delta/\gamma}$, yielding a thinner upper tail for consumption than for wealth. When $\delta=\gamma$, they lower the positive limiting propensity to consume; when $\delta>\gamma$, they become asymptotically irrelevant. The framework encompasses joy-of-giving/warm-glow bequests under random mortality. A calibrated model shows that the asymptotic characterization is accurate at observed wealth levels and reproduces wealthy households' high saving rates.

\medskip

\noindent
\textbf{Keywords:} consumption and wealth inequality, marginal propensity to consume, power law, saving rate, wealth in utility.
\end{abstract}
\medskip

\section{Introduction}

Wealthy households save a substantially larger fraction of their income than the rest of the population, and this high saving rate persists even at the very top of the wealth distribution \citep{DynanSkinnerZeldes2004,FagerengHolmMollNatvikWP,mian2025saving}. Standard optimal saving models with homothetic preferences struggle to explain this pattern: consumption is typically asymptotically linear in wealth, so the marginal propensity to consume (MPC) converges to a positive constant and saving rates flatten out \citep{Carroll1997}. This asymptotic linearity also implies that consumption inherits the tail behavior of wealth, whereas empirically consumption is markedly less concentrated than wealth in the upper tail \citep{TodaWalsh2015JPE,Lee2025,GHWW2023}. Although sufficiently strong investment risk can generate vanishing asymptotic MPCs even under homothetic preferences \citep{MaToda2021JET,MaToda2022JME}, \citet{GHWW2023} argue that the required conditions are unlikely to hold under empirically plausible calibrations.

In response, a growing quantitative literature departs from homotheticity by letting wealth enter utility directly, with period utility of the form $U(c,w)=u(c)+v(w)$. 
Such preferences date back at least to \citet{Kurz1968} and arise as the spirit of capitalism or direct preference for wealth \citep{Zou1994,Zou1995,BakshiChen1996,Carroll2002Save,Kamihigashi2008spirit}, and as joy-of-giving or warm-glow bequest motives \citep{AbelWarshawsky1988,DeNardi2004,MianStraubSufi2021}. Recent quantitative studies show that wealth-in-utility preferences---typically with wealth utility less curved than consumption utility---are key to matching the saving behavior of the rich and the upper tail of the wealth distribution \citep{straub2019consumption,GHWW2023,fella2024saving,Hubmer2026}. \citet{Tokuoka2012} provides empirical evidence consistent with this curvature ordering. Yet despite their growing empirical importance, the theoretical foundations of optimal saving with wealth preferences remain largely undeveloped: it is not known under what conditions optimal policies exist and are unique, nor how wealth preferences shape the long-run relationship between consumption and wealth.

These theoretical gaps matter for both interpretation and reliability of the quantitative results based on wealth preferences. The reason they are open is that the workhorse toolkit fails. In common specifications, the period reward function $u(c)+v(w)$ is unbounded over the state space, so the value function cannot be confined to a space of bounded functions: the Bellman operator remains monotone, but Blackwell's discounting condition breaks down. Quantitative studies therefore rely on numerical solutions, which cannot validate the underlying theory. Without a uniqueness theorem, different algorithms, such as value function iteration and the endogenous grid method, could in principle converge to different policy functions, and computational convergence alone cannot rule this out. A rigorous theoretical foundation is thus a prerequisite for the growing quantitative literature that builds on these preferences.

This paper provides that foundation and uses it to derive sharp and empirically relevant predictions. Our first contribution is to establish the existence, uniqueness, and computability of optimal consumption policies in a general incomplete-markets economy with borrowing constraints, stochastic discounting, stochastic returns, and Markov state dynamics. This framework encompasses a broad class of heterogeneous-agent models, including those with heterogeneous discount factors, idiosyncratic returns, and persistent-transitory earnings. Building on the policy iteration approach of \citet{Coleman1990} and \citet{LiStachurski2014}, we overcome the failure of standard contraction arguments caused by unbounded rewards by working in marginal utility space, where we establish a Perov contraction \citep{Perov1964,Toda2021ORL}, whose modulus is a nonnegative matrix with spectral radius below one. Theorem~\ref{thm:existence} demonstrates that the resulting time iteration operator admits a unique fixed point, that iterations converge globally at a geometric rate, and that the fixed point satisfies the transversality condition, ensuring optimality rather than merely solving the Euler equation. The associated spectral conditions generalize the familiar $\beta<1$ and $\beta R<1$ conditions in constant-coefficient models, imposing restrictions only on long-run average growth rates and allowing $\beta_t>1$ or $\beta_tR_t>1$ in individual periods.

Our second and central contribution is to provide a complete characterization of the upper-tail consumption and saving behavior through a sharp organizing principle: the relative curvature of consumption and wealth utility. Under power specifications $u(c)=c^{1-\gamma}/(1-\gamma)$ and $v(w)=\psi w^{1-\delta}/(1-\delta)$, the asymptotic MPC $\lim_{w\to\infty}c(w,z)/w$ exists and falls into three economically distinct regimes.

When $\delta<\gamma$, the asymptotic MPC is zero (Theorem~\ref{thm:MPC0}). The mechanism is fundamentally preference-based: as the marginal utility of wealth declines more slowly than the marginal utility of consumption, the incentive to preserve wealth strengthens as wealth accumulates, causing households to consume a vanishing fraction of their resources. Under mild additional conditions, we obtain a sharper characterization: consumption follows a \emph{power law}, $c(w,z)\sim \kappa(z)w^{\delta/\gamma}$ as $w\to\infty$, where the coefficient $\kappa(z)$ admits a closed-form expression. This result has three important implications. First, since $\delta/\gamma<1$, consumption has a thinner upper tail than wealth. In particular, if wealth has Pareto exponent $\zeta_w$, the implied Pareto exponent of consumption is $\zeta_w\gamma/\delta$. This provides a theoretical foundation for the empirical tail patterns documented by \citet{TodaWalsh2015JPE}, \citet{Lee2025}, and \citet{GHWW2023}. Second, the exponent $\delta/\gamma$ is directly observable as the slope of the log-log plot of consumption against wealth, providing a direct structural link between upper-tail data and preference parameters that can discipline calibration and estimation. Third, the closed-form coefficient $\kappa(z)$ provides a theoretically justified rule for extrapolating consumption policies beyond finite wealth grids, complementing numerical methods such as the Pareto extrapolation approach of \citet{Gouin-BonenfantToda2023QE}. This is particularly useful in quantitative applications where an accurate characterization of upper-tail behavior is essential.

When $\delta\ge \gamma$, wealth preferences no longer generically generate vanishing asymptotic MPCs. The asymptotic MPC remains positive (Theorems~\ref{thm:MPC2}--\ref{thm:MPC1}), except under restrictive spectral-radius conditions (Proposition~\ref{prop:MPC0}). When $\delta>\gamma$, we obtain an \emph{asymptotic irrelevance result} (Theorem~\ref{thm:MPC2}): wealth preferences may affect consumption at finite wealth levels, but their influence disappears in the upper tail, and the asymptotic MPC coincides with that of the benchmark model without wealth preferences. By contrast, when $\delta=\gamma$, wealth preferences remain active in the upper tail and determine the asymptotic MPC through a nonlinear fixed-point characterization (Theorem~\ref{thm:MPC1}). Whenever the benchmark asymptotic MPC is positive, this fixed point implies a lower limiting MPC than in the case $\delta>\gamma$. Thus, the knife-edge case $\delta=\gamma$ marks the precise boundary between the asymptotic relevance and irrelevance of wealth preferences.

This taxonomy provides a sharp guide for applied research: wealth preferences fundamentally reshape upper-tail saving behavior if and only if wealth utility is strictly less curved than consumption utility. By contrast, when $\delta\ge\gamma$, wealth preferences alone cannot generate persistent high saving rates at the top, regardless of the strength parameter $\psi$. This provides a theoretical rationale for quantitative studies that impose low curvature on wealth or bequest utility to match the saving behavior of the rich and explain wealth inequality \citep{straub2019consumption,GHWW2023,fella2024saving,Hubmer2026}. 

Our framework also encompasses one of the most common applications of wealth-in-utility preferences: joy-of-giving/warm-glow bequests. We demonstrate that a perpetual-youth model with random mortality and bequest utility is \emph{exactly} equivalent to our wealth-preference framework, with survival risk absorbed into the discount factor and mortality determining the effective weight on wealth utility. Therefore, all optimality and asymptotic results apply directly to this class of overlapping-generations models, with the comparison between $\delta$ and $\gamma$ retaining the same economic interpretation in terms of bequest curvature.

To assess the quantitative relevance of the theory at finite wealth levels, we complement the theoretical analysis with a quantitative evaluation. We calibrate an incomplete-markets model with persistent-transitory income risk, occupational switching, and idiosyncratic investment risk, using the established power-law property to calibrate $\delta/\gamma$ through the observed consumption and wealth Pareto exponents. Three findings stand out. First, relative to the classical benchmark ($\psi=0$), the wealth-preference model substantially improves the fit of the untargeted wealth distribution, reducing the implied Pareto exponent from 2.9 to 1.4, although it still understates the top 1\% wealth share. Second, the model generates a steep gradient in saving rates across the wealth distribution documented by \citet{mian2025saving}, whereas the classical benchmark implies sharply negative saving rates at the top. Third, and most importantly for applied users of our theory, convergence to the asymptotic regime is fast: the normalized consumption function $c(w,z)/w^{\delta/\gamma}$ is close to its theoretical limit by the top 1\% of the wealth distribution, so the asymptotic formulas are accurate approximations at empirically relevant wealth levels.

\paragraph{Related literature}

Our paper connects the quantitative literature on wealth preferences with the theoretical literature on income fluctuation problems. On the quantitative side, \citet{straub2019consumption}, \citet{GHWW2023}, \citet{fella2024saving}, and \citet{Hubmer2026} introduce preferences for wealth or bequests to account for the saving behavior of wealthy households, while wealth-in-utility preferences also feature prominently in studies of wealth inequality and macroeconomic policy \citep{KumhofRanciereWinant2015,MianStraubSufi2021,MichaillatSaez2021,MichauOnoSchlegl2023}. These studies differ in environments and motivations, but our theory reveals the common mechanism underlying their long-run implications---the relative curvature of consumption and wealth utility---and establishes the existence and uniqueness results that their numerical solutions presuppose.

On the theoretical side, our methods build on the fixed point theory for income fluctuation problems developed by \citet{LiStachurski2014}, \citet{MaStachurskiToda2020JET}, and \citet{MaToda2021JET,MaToda2022JME}, but incorporating wealth preferences requires new arguments and leads to new economic implications. First, wealth preferences raise saving incentives and reshape the long-run wealth dynamics, so establishing the transversality condition requires additional integrability arguments; we further show by counterexample that these conditions cannot be omitted. Second, the Perov contraction approach delivers uniqueness and globally convergent time iteration in one step, replacing the longer local-contraction arguments of earlier work. Third, and most importantly, our theory uncovers a new preference-based mechanism for vanishing MPCs, derives a precise power-law consumption behavior, and establishes a unified taxonomy of the long-run effects of wealth preferences absent in earlier studies. In the homothetic case, a zero asymptotic MPC requires restrictive spectral conditions, which under constant returns implies $R<1$ \citep{MaToda2021JET}. With wealth preferences, by contrast, the purely preference-based curvature condition $\delta<\gamma$ suffices regardless of returns and discounting.

\section{A simple two-period economy}\label{sec:simple}
Before turning to the general stochastic model, we develop the key intuition in a simple two-period model without uncertainty or labor income. The purpose is not to provide a complete characterization, but to isolate the economic force driving our main results.

An agent enters period $0$ with wealth $w_0$ and chooses consumption over two periods to solve:
\begin{subequations}\label{eq:os_simple}
\begin{align}
    &\maximize && U(c_0,w_0)+\beta U(c_1,w_1) \label{eq:U_simple} \\
  &\st
    && c_0\le w_0 \quad \text{and} \quad c_1\le w_1=R(w_0-c_0), \label{eq:budget_simple}
\end{align}
\end{subequations}
where $c_t,w_t>0$ denote consumption and wealth in period $t$, and $\beta, R >0$ are the discount factor and gross return on wealth, respectively. We use the power utility function defined by
\begin{equation}
    p(c;\gamma)=\begin{cases*}
        \frac{c^{1-\gamma}}{1-\gamma} & if $0<\gamma$ and $\gamma \neq 1$,\\
        \log c & if $\gamma=1$,
    \end{cases*}
    \label{eq:CRRA}
\end{equation}
where $\gamma>0$ is the coefficient of relative risk aversion. Suppose the consumption and wealth utility functions are respectively $u(c)=p(c;\gamma)$ and $v(w)=\psi p(w;\delta)$ for some $\gamma,\delta,\psi>0$. 

Letting $w=w_0$, the Euler equation characterizing optimal consumption is
\begin{equation}
    c^{-\gamma} = \beta R([R(w-c)]^{-\gamma}+\psi[R(w-c)]^{-\delta}). \label{eq:foc_simple}
\end{equation}
The asymptotic behavior of consumption is determined by the relative importance of the marginal utilities of consumption and wealth in this equation. The following proposition summarizes the asymptotic implications of this simple environment.

\begin{prop}\label{prop:c_simple}
The optimal saving problem \eqref{eq:os_simple} has a unique optimal policy $c(w)$ that satisfies \eqref{eq:foc_simple}. Moreover, the following statements hold:
\begin{enumerate}
    \item\label{item:c_simple0} If $\delta<\gamma$, then the optimal policy obeys a power law, that is,
    \begin{equation*}
        \lim_{w\to\infty}\frac{c(w)}{w}=0 
        \quad \text{and} \quad 
        \lim_{w\to\infty}\frac{c(w)}{w^{\delta/\gamma}}=(\psi \beta R^{1-\delta})^{-1/\gamma}.
    \end{equation*}
    \item\label{item:c_simple1} If $\delta=\gamma$, then $\lim_{w\to\infty}c(w)/w=\bar{c}_1$, where $\bar{c}_1 \in (0,1)$ is the unique constant such that $c(w)=\bar c_1 w$ satisfies \eqref{eq:foc_simple}.
    \item\label{item:c_simple2} If $\delta>\gamma$, then $\lim_{w\to\infty}c(w)/w=\bar{c}_2$, where $\bar{c}_2= 1/(1+(\beta R^{1-\gamma})^{1/\gamma}) >\bar{c}_1$.
\end{enumerate}
\end{prop}

We prove the proposition in the main text because it is useful for developing intuition.

\begin{proof}
For fixed $w>0$, define $f(c)\coloneqq \beta R([R(w-c)]^{-\gamma}+\psi[R(w-c)]^{-\delta})-c^{-\gamma}$ for $c\in (0,w)$. The function $f$ is continuous and strictly increasing, with $f(c)\to-\infty$ as $c\downarrow0$ and $f(c)\to\infty$ as $c\uparrow w$. Hence, there exists a unique
$c(w)\in(0,w)$ satisfying \eqref{eq:foc_simple}. Strict concavity
of the objective implies that this solution is the unique optimum.

\medskip
\noindent \ref{item:c_simple0}
Since \eqref{eq:foc_simple} implies
\begin{equation}
    c^{-\gamma}\ge \beta R(\psi[R(w-c)]^{-\delta})
    \ge \psi\beta R^{1-\delta}w^{-\delta},
    \label{eq:c_ub_power1}
\end{equation}
simple algebra yields
\begin{equation}
    0<\frac{c(w)}{w}\le (\psi\beta R^{1-\delta})^{-1/\gamma}w^{\delta/\gamma-1}\to 0 \label{eq:MPC_ub}
\end{equation}
as $w\to\infty$. Moreover, for any $\epsilon\in (0,1)$, it follows from \eqref{eq:MPC_ub} that $c(w)\le \epsilon w$ for large enough $w$. Then \eqref{eq:foc_simple} implies
\begin{align*}
    c^{-\gamma}
    &\le \beta R^{1-\gamma}(1-\epsilon)^{-\gamma} w^{-\gamma}+\psi\beta R^{1-\delta}(1-\epsilon)^{-\delta} w^{-\delta} \\
    \implies w^{-\delta/\gamma} c&\ge (\beta R^{1-\gamma}(1-\epsilon)^{-\gamma} w^{\delta-\gamma}+\psi\beta R^{1-\delta}(1-\epsilon)^{-\delta})^{-1/\gamma}.
\end{align*}
Letting $w\uparrow\infty$ and then $\epsilon\downarrow 0$, we obtain $\liminf_{w\to\infty} w^{-\delta/\gamma}c(w)\ge (\psi\beta R^{1-\delta})^{-1/\gamma}$. But \eqref{eq:c_ub_power1} also implies $\limsup_{w\to\infty} w^{-\delta/\gamma}c(w)\le (\psi\beta R^{1-\delta})^{-1/\gamma}$, so in fact we have
\begin{equation}
    \lim_{w\to\infty} w^{-\delta/\gamma}c(w)=(\psi\beta R^{1-\delta})^{-1/\gamma}. \label{eq:c_powerlaw_1}
\end{equation}
That is, the consumption function behaves like the power function $w^{\delta/\gamma}$.

\medskip
\noindent \ref{item:c_simple1}
Since $\delta=\gamma$, dividing \eqref{eq:foc_simple} by $w^{-\gamma}$ and setting $\bar{c}_1=c/w$ yields
\begin{equation}
    \bar{c}_1^{-\gamma}=\beta R([R(1-\bar{c}_1)]^{-\gamma}+\psi[R(1-\bar{c}_1)]^{-\gamma}). \label{eq:foc_cbar1}
\end{equation}
By the same reasoning as in case \ref{item:c_simple0}, equation \eqref{eq:foc_cbar1} has a unique solution $\bar{c}_1\in (0,1)$. Therefore, the consumption function is linear, $c(w)=\bar{c}_1w$.

\medskip
\noindent \ref{item:c_simple2} Note that \eqref{eq:foc_simple} implies
\begin{equation}
    \beta R[R(w-c)]^{-\gamma}\le c^{-\gamma}=\beta R([R(w-c)]^{-\gamma}+\psi[R(w-c)]^{-\delta}). \label{eq:foc_ineq}
\end{equation}
Solving the left inequality of \eqref{eq:foc_ineq} for $c$ yields $c(w)\le \bar{c}_2w$, where $\bar{c}_2\in (0,1)$ solves
\begin{equation}
    \bar{c}_2^{-\gamma}=\beta R[R(1-\bar{c}_2)]^{-\gamma}\iff \bar{c}_2=\frac{1}{1+(\beta R^{1-\gamma})^{1/\gamma}}. \label{eq:foc_cbar2}
\end{equation}
Dividing \eqref{eq:foc_ineq} by $w^{-\gamma}$ and setting $m(w)\coloneqq c(w)/w\le \bar{c}_2$, we obtain
\begin{equation}
    m^{-\gamma}\le \beta R^{1-\gamma}(1-m)^{-\gamma}+\psi\beta R^{1-\delta}w^{\gamma-\delta}(1-\bar{c}_2)^{-\delta}. \label{eq:foc_ineq2}
\end{equation}
Since $\delta>\gamma$, the last term in \eqref{eq:foc_ineq2} converges to 0 as $w\to \infty$. Noting that $m\le \bar{c}_2$ is bounded above, for any $\epsilon\in (0,1)$, for sufficiently large $w$ we have
\begin{equation*}
    m^{-\gamma}\le \beta R^{1-\gamma}(1-m)^{-\gamma}+\epsilon m^{-\gamma}\iff \frac{c(w)}{w}=m(w)\ge \frac{1}{1+(\beta R^{1-\gamma}/(1-\epsilon))^{1/\gamma}}.
\end{equation*}
Letting $w\uparrow\infty$ and then $\epsilon\downarrow 0$, it follows that $\lim_{w\to\infty}c(w)/w=\bar{c}_2$. Thus, the consumption function is asymptotically linear with an asymptotic slope of $\bar{c}_2>\bar{c}_1$.
\end{proof}

Proposition~\ref{prop:c_simple} highlights three distinct regimes.
\begin{enumerate*}
    \item When $\delta<\gamma$, the marginal utility of wealth declines more slowly than the marginal utility of consumption as wealth increases. Therefore, the incentive to preserve wealth becomes increasingly important in the upper tail. Consumption must grow more slowly than wealth to satisfy the Euler equation. As a result, the consumption-wealth ratio converges to zero, and consumption follows a power-law relationship with wealth, with exponent $\delta/\gamma$.
    \item When $\delta=\gamma$, consumption and wealth utility have identical curvature. The two sources of utility therefore remain balanced as wealth grows. The consumption-wealth ratio converges to a positive constant that depends on both consumption and wealth preferences. In this case, scaling consumption by wealth eliminates the effect of wealth levels asymptotically.
    \item Finally, when $\delta>\gamma$, the marginal utility of wealth declines faster than the marginal utility of consumption. Although wealth preferences continue to affect choices at finite wealth levels, their effect on the consumption-wealth ratio becomes asymptotically negligible. Consequently, the upper-tail consumption behavior converges to that of the benchmark model without wealth utility.
\end{enumerate*}

Despite its simplicity, this two-period benchmark already reveals the central mechanism of the paper: the relative curvature of consumption and wealth utility governs the asymptotic saving behavior and determines whether wealthy households continue to accumulate assets or gradually consume them. The rest of the paper shows that this insight survives in a fully general optimal saving framework. 

\section{Optimal saving problem}
\label{sec:os}

We now embed the wealth-preference mechanism into a general infinite-horizon optimal saving framework and establish existence, uniqueness, and computability of the optimal consumption policy. These results provide the foundation for our analysis of upper-tail consumption and saving behavior in Section~\ref{sec:asymptotic}.

\subsection{Economic environment}
\label{subsec:os_problem}

We consider a discrete-time infinite-horizon economy in which the household derives utility from both current consumption and wealth. The household faces stochastic returns, stochastic discounting, labor income risk, and borrowing constraints, and chooses consumption and wealth $\set{(c_t, w_t)}_{t=0}^\infty$ to maximize expected lifetime utility:
\begin{subequations}\label{eq:os}
\begin{align}
    &\maximize && \E_0 \set{\sum_{t = 0}^\infty \left(\prod_{i=1}^t \beta_i \right) [u(c_t) + v(w_t)]} \label{eq:os_obj} \\
  &\st
    && w_{t+1} = R_{t+1} (w_t - c_t) + Y_{t+1}, \label{eq:os_motion} \\
    &&& 0 \le c_t \le w_t, \label{eq:os_budget}
\end{align}
\end{subequations}
where the initial state $(w_0,Z_0)=(w,z)$ is given. Here $u,v$ denote the utility from consumption and wealth, respectively; $\beta_t\ge 0$ is the discount factor; $R_t\ge 0$ is the gross rate of return on wealth; and $Y_t$ denotes nonfinancial income, which is strictly positive almost surely. We make the following assumption.

\begin{asmp}[Preferences]\label{asmp:U}
The utility functions $u,v:(0,\infty)\to \R$ are continuously differentiable, $u'$ is strictly decreasing and satisfies $u'>0$ and $u'(0)=\infty$, and $v'$ is decreasing and satisfies $v'\ge 0$.
\end{asmp}

Furthermore, the stochastic process $\set{(\beta_t,R_t,Y_t)}_{t=1}^\infty$ evolves according to
\begin{equation}
    \beta_t = \beta(Z_{t-1},Z_t,\epsilon_t), \quad
    R_t = R(Z_{t-1},Z_t,\epsilon_t), \quad \text{and} \quad
    Y_t = Y(Z_{t-1},Z_t, \epsilon_t), \label{eq:RY_func}
\end{equation}
where $\beta,R,Y$ are measurable nonnegative functions, $\set{Z_t}$ is a finite-state Markov chain, and $\set{\epsilon_t}$ is a sequence of \iid shocks independent of $\set{Z_t}$. This formulation allows the discount factor, return, and income to depend on both the current and lagged Markov states together with an \iid\ innovation. It is intentionally flexible: both $Z_t$ and $\epsilon_t$ may be vector-valued, and $\epsilon_t$ may have either discrete or continuous support, allowing the framework to encompass a broad class of heterogeneous-agent models studied in quantitative macroeconomics. The model in Section~\ref{sec:quant}, with occupational switching and a persistent-transitory income process, falls into this category.

Throughout, conditional probabilities and expectations are denoted by
\begin{align*}
    \Pr_z&\coloneqq \Pr[\,\cdot \mid Z_0=z], & \E_z&\coloneqq \E[ \, \cdot \mid Z_0 = z], & \E_{w,z}&\coloneqq \E[ \,\cdot \mid (w_0,Z_0) = (w, z)],
\end{align*}
with analogous notation used for other conditioning events. We denote the next period value of a random variable $X$ by $\hat{X}$. For instance, we write $\hat{R}=R(Z,\hat{Z},\hat{\epsilon})$ for $(Z_t,Z_{t+1},\epsilon_{t+1})=(Z,\hat{Z},\hat{\epsilon})$ to simplify notation.

\subsection{Optimality conditions}\label{ss:opt_conds}
Since $Y_t>0$ almost surely, the law of motion \eqref{eq:os_motion} implies that $w_t>0$ almost surely for all $t$. Therefore, without loss of generality, we take the wealth space to be $(0,\infty)$. Letting $\ZZ$ denote the state space for $\set{Z_t}$, the state space for $\set{(w_t, Z_t)}_{t \ge 0}$ is then $\SS\coloneqq (0, \infty) \times \ZZ$. A \emph{feasible policy} is a Borel measurable function $c \colon 
\SS \to \R$ with $0 < c(w,z) \le w$ for all $(w,z) \in \SS$. A 
feasible policy $c$ and initial condition $(w,z) \in \SS$ generate a wealth path $\set{w_t}_{t \ge 0}$ via \eqref{eq:os_motion} when $c_t = c(w_t, Z_t)$ and $(w_0, Z_0) = (w,z)$.

Because we do not assume that the period utility functions $u,v$ are bounded or the discount factor $\beta_t$ is less than 1, the lifetime utility \eqref{eq:os_obj} may not be well defined. We therefore define optimality by the overtaking criterion \citep{Brock1970}. Given a feasible policy $c$, let $(u \circ c + v)(w,z)=u(c(w,z))+v(w)$ and
\begin{equation}
    V_{c,T}(w,z)\coloneqq \E_{w,z} \sum_{t = 0}^T 
        \left(\prod_{i=1}^t \beta_i\right) 
        (u \circ c + v) (w_t, Z_t) \label{eq:Vc}
\end{equation}
be the utility up to time $T$ conditional on $(w_0,Z_0)=(w,z)$, where $\set{w_t}$ is the wealth path generated by $(c,(w,z))$. For two feasible policies $c_1,c_2$, we say that $c_1$ \emph{overtakes} $c_2$ if
\begin{equation*}
    \limsup_{T\to\infty}(V_{c_2,T}(w,z)-V_{c_1,T}(w,z))\le 0
\end{equation*}
for all $(w,z)\in \SS$. We say that a feasible policy $c^*$ is \emph{optimal} if it overtakes any other feasible policy $c$.

In Appendix \ref{ss:foc_tvc}, we show that the first-order optimality condition for the problem can be written as
\begin{equation}
    \left(u' \circ c\right)(w,z) = 
    \max \set{\E_z \hat{\beta} \hat{R}(u' \circ c + v') (\hat{w}, \hat{Z}), u'(w)} \label{eq:foc}
\end{equation}
for all $(w,z) \in \SS$, which we refer to as the \emph{Euler equation}. Intuitively, the marginal utility of current consumption equals the discounted marginal value of transferring resources to the future through saving. The latter includes not only the discounted marginal utility from future consumption, as in standard saving models, but also the marginal utility of wealth generated by future asset holdings through $v'$. The max operator captures the borrowing constraint: when households would like to consume more than their available resources, they are constrained to consume all current wealth, so that $c(w,z)=w$. 

Moreover, an optimal policy must satisfy the \emph{transversality condition}:
\begin{equation}
    \lim_{t \to \infty}     
    \E_{w,z} \left(\prod_{i=1}^t \beta_i\right) 
    \left(u' \circ c \right)(w_t, Z_t)(w_t - c_t) = 0 \label{eq:tvc}
\end{equation}
for all $(w, z) \in \SS$. This condition rules out optimal paths along which households retain wealth whose discounted marginal value remains positive forever. 

We summarize these optimality conditions in the following proposition, which underlies our subsequent analysis of the optimal consumption function. The proof follows directly from \citet[Proposition 15.2 and Theorem 15.3]{TodaEME} and is therefore omitted.

\begin{prop}[Sufficiency of first-order and transversality conditions]
    \label{prop:optimal}
    If Assumption~\ref{asmp:U} holds, any feasible policy $c$ satisfying the Euler equation \eqref{eq:foc} and the transversality condition \eqref{eq:tvc} is optimal.
\end{prop}

\subsection{Existence and computability of optimal consumption}

We now construct the optimal consumption policy as a fixed point.

\subsubsection{Time iteration approach beyond standard contraction}

A key difficulty in solving the optimal saving problem \eqref{eq:os} is that the utility functions $u$ and $v$ are unbounded in common specifications such as $u(c)=\log c$. Hence, the value function cannot be confined to a space of bounded functions. While the associated Bellman operator remains monotone, it fails to satisfy Blackwell's discounting condition. As a result, standard contraction mapping arguments based on value function iteration over bounded function spaces are not applicable.

To overcome this challenge, we build on \citet{LiStachurski2014} and shift attention from value functions to policy functions. Working in a space of marginal utilities, we establish a Perov contraction, which yields existence and uniqueness of a policy satisfying the Euler equation.

Let $\cC$ be the space of continuous functions $c \colon \SS \to \R$ such that $c$ is increasing in the first argument, $0 < c(w,z) \le w$ for all $(w,z) \in \SS$, and
\begin{equation}
    \label{eq:C4}
   \sup_{(w,z) \in \SS} \abs{(u' \circ c)(w,z) - u'(w)} < \infty.
\end{equation}
To compare two consumption policies, we pair $\cC$ with the distance
\begin{equation}
\label{eq:rho_metric}
  \rho(c_1,c_2) 
    \coloneqq \norm{u' \circ c_1 - u' \circ c_2}
    \coloneqq \sup_{(w,z) \in \SS} \abs{(u' \circ c_1)(w,z) - (u' \circ c_2)(w,z)},
\end{equation}
which evaluates the maximal difference in terms of marginal utility. While elements of $\cC$ are not generally bounded, $\rho$ is a
valid metric on $\cC$. In particular, $\rho$ is finite on $\cC$ since $\rho(c_1,c_2) \le \norm{u' \circ c_1 - u' } + \norm{u' \circ c_2 - u'}$, and the last two terms are finite by \eqref{eq:C4}. The resulting metric space $(\cC, \rho)$ is complete (Lemma~\ref{lem:metric}).

We aim to characterize the optimal policy as the fixed point of the \emph{time iteration operator} $T$ defined as follows: for fixed $c \in \cC$ and $(w,z) \in \SS$, the value of the image $Tc$ at $(w,z)$ is defined as the $\xi \in (0,w]$ that solves
\begin{equation}
    u'(\xi) = \psi_c(\xi, w,z) \coloneqq \max \set{g_c(\xi, w, z), u'(w)}, \label{eq:T_opr}
\end{equation}
where
\begin{equation}
    g_c(\xi,w,z) 
    \coloneqq \E_z \hat{\beta} \hat{R}(u' \circ c + v')\left(\hat{R}(w - \xi) + \hat{Y}, \hat{Z}\right) \label{eq:keypart_T_opr}
\end{equation}
is defined on the set
\begin{equation}
    G \coloneqq \set{(\xi, w, z) \in \R_+ \times (0, \infty) \times \ZZ \colon 0 < \xi \le w}. \label{eq:dom_T_opr}
\end{equation}
The time iteration operator $T$ was first introduced by \citet{Coleman1990} in the context of the stochastic growth model.

\subsubsection{Spectral radius condition and economic interpretation}
\label{sss:interp_spect}

Before introducing our main assumption, we define the key objects and interpret them economically. For $\theta\in\R$, we define the matrix $K(\theta)$ by
\begin{equation}
	K_{z\hat{z}}(\theta) \coloneqq P(z,\hat{z}) \int \beta(z, \hat{z}, \hat{\epsilon}) R(z,\hat{z},\hat{\epsilon})^\theta \pi (\diff \hat{\epsilon}), 
	\qquad (z, \hat{z} \in \ZZ)
	\label{eq:Kmat}
\end{equation}
where $P(z,\hat z)$ is the probability of transitioning from $z$ to $\hat z$ in one period, and $\pi$ is the probability distribution of $\set{\epsilon_t}$.\footnote{\label{fn:K}The matrix $K(\theta)$ is expressed as a function on $\ZZ \times \ZZ$ in \eqref{eq:Kmat} but can be represented in traditional matrix notation by enumerating $\ZZ$. In what follows, we use $K(\theta)$ for $\theta=1-\gamma$ with $\gamma\ge 0$. We adopt the conventions $R^{1-\gamma}=R\cdot R^{-\gamma}$, $0\cdot \infty=0$, and $0^0=1$ so that entries of $K(1-\gamma)$ are always well defined in $[0,\infty]$.} For a square matrix $A$, we denote its spectral radius by
\begin{equation}
	r(A) \coloneqq \max \set{\abs{\alpha}: \text{$\alpha$ is an eigenvalue of $A$}}. \label{eq:spectral_radius}
\end{equation}
In other words, $r(A)$ is the largest absolute value of all its eigenvalues.

In what follows, we frequently work with $r(K(\theta))$. As shown in Lemma~\ref{lem:Lphi}, for irreducible $\set{Z_t}$, we have
\begin{equation}\label{eq:rK}
	r(K(\theta)) = \lim_{n \to \infty} \left(
	\E \prod_{t=1}^n \beta_t R_t^\theta 
	\right)^{1/n},
\end{equation}
where $\E$ denotes expectation with respect to the unique stationary distribution of $\set{Z_t}$. Therefore, $\log r(K(\theta))$ represents the asymptotic per-period log growth rate of the expected product $G_n\coloneqq \E \prod_{t=1}^n \beta_t R_t^\theta$.\footnote{To see this,
	note that $G_n = G_0 \prod_{t=1}^n (G_t/G_{t-1})$, which implies
	\begin{equation*}
		\log r(K(\theta)) = \lim_{n\to \infty} \frac{1}{n} \log G_n
		= \lim_{n\to \infty} \frac{1}{n} \sum_{t=1}^n \log 
		\left(\frac{G_t}{G_{t-1}} \right),
	\end{equation*}
	which corresponds to the long-run average of the incremental log growth rates of $G_n$.}  

Specifically, the condition $r(K(\theta)) < 1$ implies that this growth rate is negative. Equivalently, the one-period growth factors $G_n/G_{n-1}$ are, on average, below one in the long run, so that discounting dominates return growth over time. Notably, this condition does not preclude $\beta_t R_t^\theta > 1$ in any given period.

The economic interpretation of $r(K(\theta))$ depends on the value of $\theta$. When $\theta=0$, it governs the growth rate of the cumulative discount factor. When $\theta=1$, it characterizes the growth rate of discounted returns, which in turn determines the evolution of wealth. When $\theta=1-\gamma$, it corresponds to the growth rate of discounted marginal utility under CRRA preferences, and thus plays a central role in the Euler equation. The following example clarifies the interpretation.

\begin{exmp}[Constant discount factor and returns]\label{exmp:const_beta_R}
If $\beta_t \equiv \beta$ and $R_t\equiv R$, we have $r(K(0)) = \beta$, $r(K(1)) = \beta R$, and $r(K(1-\gamma)) = \beta R^{1-\gamma}$. As a result, the condition $r(K(\theta)) < 1$ for $\theta \in \set{0,1,1-\gamma}$ reduces to the classical assumptions $\beta<1$, $\beta R<1$, and $\beta R^{1-\gamma} < 1$ in the standard income fluctuation problem \citep{LiStachurski2014}. In the terminology of \citet{Carroll2004}, the conditions $\beta R<1$ and $\beta R^{1-\gamma}<1$ correspond to the absolute and return impatience conditions.

In contrast to these classical uniform stability conditions, which impose pointwise restrictions in every period, the requirement $r(K(\theta)) < 1$ only restricts the long-run average growth rate. As a result, the model allows for $\beta_t > 1$, $\beta_t R_t > 1$, or $\beta_t R_t^{1-\gamma} > 1$ in individual periods, as long as such episodes are offset so that long-run growth remains bounded.
\end{exmp}

\subsubsection{Main assumptions and optimality result}

The preceding discussion highlights that the key requirement is a balance between discounting and return growth in the long run. We now formalize this idea in the following assumption, together with some standard integrability conditions. These conditions ensure that the Euler equation is well defined and that the time iteration operator admits a well-behaved fixed point.

\begin{asmp}[Integrability and spectral conditions]\label{asmp:Euler}
Let $K(\theta)$ be as in \eqref{eq:Kmat}.
\begin{enumerate}
    \item\label{item:Euler1} $\E_z \hat{\beta}\hat{R}<\infty$, $\E_z \hat{\beta}\hat{R}u'(\hat{Y})<\infty$, and $\E_z \hat{\beta}\hat{R}v'(\hat{Y})<\infty$ for all $z\in \ZZ$ and $r(K(1))<1$.
    \item\label{item:Euler2} $\E_z\hat{\beta}\hat{Y}<\infty$ for all $z\in \ZZ$ and $r(K(0))<1$.
\end{enumerate}
\end{asmp}

Conditions \ref{item:Euler1} and \ref{item:Euler2} impose spectral radius restrictions ensuring that discounting dominates return growth in the long run, replacing the classical requirements $\beta R < 1$ and $\beta<1$.

The following theorem shows that Assumptions~\ref{asmp:U} and \ref{asmp:Euler} imply the existence and uniqueness of an optimal policy.

\begin{thm}[Existence, uniqueness, and computability of optimal policies]\label{thm:existence}
If Assumptions~\ref{asmp:U} and \ref{asmp:Euler} hold, then the following statements hold.
\begin{enumerate}
    \item\label{item:existence1} $T:\cC\to \cC$ is a monotone self-map and has a unique fixed point $c^* \in \cC$.
    \item\label{item:existence2} For all $c \in \cC$, we have $\rho(T^n c, c^*) \to 0$ as $n \to \infty$. Furthermore,
    \begin{equation*}
        \limsup_{n\to\infty}\rho(T^nc,c^*)^{1/n}\le r(K(1)).
    \end{equation*}
    \item\label{item:existence3} The transversality condition \eqref{eq:tvc} holds and $c^*$ is an optimal policy.
\end{enumerate}
\end{thm}

Part~\ref{item:existence2} shows that, under our conditions, the familiar time iteration algorithm is globally convergent and bounds its convergence rate. In what follows, we refer to $c^*$ in Theorem~\ref{thm:existence} as the \emph{optimal consumption function}.

\begin{rem}[Necessity of additional integrability restrictions for transversality]
As shown in the proof of Theorem~\ref{thm:existence}, the existence, uniqueness, and convergence results in parts~\ref{item:existence1} and~\ref{item:existence2} do not require Assumption~\ref{asmp:Euler}\ref{item:Euler2}. But an additional condition such as Assumption~\ref{asmp:Euler}\ref{item:Euler2} is needed to establish the transversality condition. To see this, consider the following example. 

Let $u(c)=\log c$ and $v(w)=w+\log w$; let $\beta, R$ be constant with $\beta R\in (0,1)$; and let income $\set{Y_t}$ be \iid with $Y_t\ge 1$ and $\E[Y_t]=\infty$. Since $u'(c)=1/c$ and $v'(w)=1+1/w$, Assumption \ref{asmp:U} clearly holds. Since $\hat{Y}\ge 1$ and $r(K(1))=\beta R<1$, Assumption \ref{asmp:Euler}\ref{item:Euler1} holds but \ref{item:Euler2} fails. Since $v'(w)\ge 1$, the Euler equation \eqref{eq:foc} implies $1/c_t\ge \beta R$, so $c_t\le 1/(\beta R)$. Therefore, saving satisfies $s_t=w_t-c_t\ge Y_t-1 /(\beta R)$. Consequently,
\begin{equation*}
    \E[\beta^tu'(c_t)s_t]\ge \beta^t(\beta R)\E[\max\set{0,Y_t-1/(\beta R)}]\ge \beta^{t+1}R\E[Y_t-1/(\beta R)]=\infty
\end{equation*}
and the transversality condition \eqref{eq:tvc} fails.
\end{rem}

\subsection{Elementary properties of consumption and saving}

This section studies the properties of the optimal consumption function obtained in Theorem~\ref{thm:existence}. We maintain Assumptions~\ref{asmp:U} and \ref{asmp:Euler} throughout. Let $s^*(w,z) \coloneqq w - c^*(w,z)$ denote the optimal saving function. The next two propositions establish monotonicity with respect to wealth, income, and the marginal utility of wealth.

\begin{prop}[Monotonicity with respect to wealth]\label{prop:monotonew}
The optimal consumption and saving functions $c^*(w,z)$ and $s^*(w,z)$ are increasing in $w$.
\end{prop}

\begin{prop}[Monotonicity with respect to income and marginal utility of wealth]\label{prop:monotone_Yv}
For $j\in\set{1,2}$, let $c_j^*$ be the optimal consumption function corresponding to income process $Y_{jt}$ and wealth utility $v_j$, holding other primitives fixed. If there exists $m\ge 0$ such that $m\le Y_{1t}\le Y_{2t}$ almost surely and $v_1'(w)\ge v_2'(w)$ for $w\in[m,\infty)\cap (0,\infty)$, then $c_1^* \le c_2^*$ pointwise on $\SS$.\footnote{The proposition nests two useful special cases. By setting $m=0$ and $v_1=v_2$, it delivers monotonicity with respect to income alone, while by setting $Y_{1t}=Y_{2t}$, it delivers monotonicity with respect to the marginal utility of wealth alone.}
\end{prop}

The next proposition characterizes the borrowing constraint in terms of a state-dependent threshold: agents optimally exhaust their resources if and only if wealth falls below a critical level.

\begin{prop}[Threshold for saving decision]\label{prop:binding}
Letting
\begin{equation}
    \bar{w}(z) \coloneqq (u')^{-1}\left(\max\set{\E_z \hat{\beta} \hat{R} (u' \circ c^* + v') (\hat{Y}, \hat{Z}),u'(\infty)}\right)\in (0,\infty], \label{eq:w_bar}
\end{equation}
we have $c^*(w,z) = w$ if and only if $w \le \bar{w}(z)$.
\end{prop}

Proposition~\ref{prop:binding} establishes that, for each state $z$, there exists a threshold $\bar w(z)$ below which the borrowing constraint binds and the agent consumes all available resources. This threshold reflects the trade-off between the high marginal utility of current consumption at low wealth levels and the continuation value of saving, which in our framework is driven by both precautionary motives and the direct marginal utility from wealth $v'(w)$. A stronger preference for wealth raises the marginal value of saving and therefore lowers the threshold $\bar w(z)$, making agents less likely to exhaust their resources even when wealth is low.\footnote{This result complements \citet[Proposition~A.1]{achdou2022income},
who show that high-income households may save even at the asset lower bound
in a continuous-time Aiyagari model. The difference lies in the state variable:
their state is financial assets, whereas our $w_t$ denotes total available
resources, including current income. Thus, high income at the asset lower bound
corresponds to high $w_t$ in our model, so the two results are consistent.}

In the classical case $v\equiv 0$, further properties such as concavity of the consumption function are known \citep{CarrollKimball1996,MaStachurskiToda2020JET,CarrollHolmKimball2021,Toda2021JME}; we leave such extensions for future research.

\section{Asymptotic properties of consumption}\label{sec:asymptotic}

In this section, we characterize the asymptotic behavior of optimal consumption at high wealth levels. The analysis reveals how the relative curvature of consumption and wealth utility shapes upper-tail consumption and saving behavior. To this end, we introduce the following assumption.

\begin{asmp}[Power utility]\label{asmp:power_u}
The period utility functions satisfy $u(c)=p(c;\gamma)$ and $v(w)=\psi p(w;\delta)$ for some $\gamma,\delta,\psi>0$, where $p$ is given by \eqref{eq:CRRA}.
\end{asmp}

We organize the analysis around three cases: $\delta<\gamma$, $\delta=\gamma$, and $\delta>\gamma$, which imply distinct asymptotic properties of the consumption function. We first consider the case $\delta<\gamma$, where the asymptotic MPC vanishes and, under additional conditions, consumption exhibits a power-law behavior.

\subsection{Asymptotic MPCs for \texorpdfstring{$\delta<\gamma$}{}}\label{ss:0ampc}

Before turning to the role of wealth preferences, we examine whether the existing stochastic mechanism for generating vanishing MPCs remains valid in our framework. The following proposition extends the zero asymptotic MPC result of \citet[Theorem~3]{MaToda2021JET} to economies with wealth preferences.

\begin{prop}[Zero asymptotic MPC under restrictive spectral conditions]\label{prop:MPC0}
If Assumptions~\ref{asmp:U}--\ref{asmp:power_u} hold, $K(1-\gamma)$ is irreducible,\footnote{A square matrix $A$ is \emph{reducible} if there exists a permutation matrix $M$ such that $M'AM$ is block upper triangular with diagonal blocks. Otherwise, $A$ is called \emph{irreducible}.} and $r(K(1-\gamma))\ge 1$, then for all $z\in \ZZ$, we have
\begin{equation}
	\lim_{w\to \infty} \frac{c^*(w,z)}{w} = 0. \label{eq:MPC0}
\end{equation}
\end{prop}

\begin{rem}
Proposition~\ref{prop:MPC0} shows that the stochastic mechanism identified by \citet{MaToda2021JET} continues to operate when wealth preferences are introduced. The same proof also covers $v\equiv 0$. However, generating a vanishing asymptotic MPC through this mechanism still requires the restrictive spectral radius condition $r(K(1-\gamma)) \ge 1$, which places joint restrictions on patience, returns, and preference parameters in view of \eqref{eq:Kmat}--\eqref{eq:rK}.
\end{rem}

We now establish our main theoretical result: wealth preferences generate vanishing MPCs through a mechanism fundamentally different from the stochastic one in Proposition~\ref{prop:MPC0}. The following theorem characterizes the asymptotic MPC and the growth rate of consumption relative to wealth under the wealth-preference mechanism.

\begin{thm}[Wealth preferences, vanishing MPC, and power-law consumption]\label{thm:MPC0}
	Suppose Assumptions~\ref{asmp:U}--\ref{asmp:power_u} hold. If $\delta<\gamma$ and $\Pr_z(\hat{\beta}\hat{R}>0)>0$ for all $z\in \ZZ$, then
	\begin{equation}
		\lim_{w\to \infty} \frac{c^*(w,z)}{w} = 0 
		\quad \text{and} \quad 
		\limsup_{w\to \infty} \frac{c^*(w,z)}{w^{\delta/\gamma}} < \infty
		\label{eq:MPC0_and_speed_conv}
	\end{equation}
    for all $z\in \ZZ$. If in addition $K(1-\gamma), K(1-\delta)$ are irreducible with $r(K(1-\gamma)), r(K(1-\delta))< 1$ and there exists $m>0$ such that $Y_t \ge m$ almost surely, then
	\begin{equation}\label{eq:speed_conv}
		\lim_{w\to \infty} \frac{c^*(w,z)}{w^{\delta/\gamma}} = v^*(z)^{-1/\gamma} \in (0,\infty) 
	\end{equation}
    for all $z\in \ZZ$, where $v^* = (v^*(z))_{z\in \ZZ} = [I - K(1-\delta)]^{-1} K(1-\delta) \psi 1$.
\end{thm}
\medskip 

\begin{rem}[Wealth preferences as a direct source of vanishing MPCs]
Theorem~\ref{thm:MPC0} demonstrates that vanishing asymptotic MPCs arise whenever two additional mild conditions are satisfied: 
\begin{enumerate*}
    \item the curvature of wealth utility is lower than that of consumption utility, $\delta<\gamma$, and
    \item the effective discounted return is positive with positive probability, $\Pr_z(\hat \beta \hat R>0)>0$ for all $z\in \ZZ$.
\end{enumerate*}
The latter is a nondegeneracy requirement ensuring that future wealth retains positive value, and is satisfied in virtually all applications of interest.

Unlike the stochastic mechanism identified by \citet{MaToda2021JET}, the wealth-preference mechanism does not require restrictive joint conditions on returns, discounting, or risk preferences. Once $\delta<\gamma$ holds, the asymptotic MPC vanishes \emph{regardless of the magnitude of wealth returns, stochastic discounting, or the specific stochastic processes governing them}, provided that the effective discounted return condition is satisfied. Therefore, wealth preferences provide a simple and robust explanation for low MPCs and high saving rates among wealthy households.
\end{rem}

\begin{rem}[Constant return rates]
    The difference between the two mechanisms is particularly transparent in a deterministic return environment. Proposition~8 of \citet{MaToda2021JET} shows that, when $v\equiv0$ and $R_t\equiv R$, vanishing asymptotic MPCs require $R<1$, a condition that is difficult to reconcile with empirically plausible returns.
    By contrast, the wealth-preference mechanism operates through the curvature of preferences rather than return dynamics. The simple condition $\delta<\gamma$ is sufficient to generate vanishing asymptotic MPCs, even in deterministic environments with $R\ge 1$.
\end{rem}

\begin{rem}[Power-law consumption and extrapolation]
Under the additional conditions in Theorem~\ref{thm:MPC0}, consumption satisfies
\begin{equation}
    c^*(w,z)\sim \kappa(z)w^{\delta/\gamma}, \label{eq:consumption_power_law}
\end{equation}
where $\kappa(z) = v^*(z)^{-1/\gamma}$ admits a closed-form characterization. Because $\delta/\gamma<1$, consumption grows more slowly than wealth and therefore has a thinner upper tail, consistent with the empirical patterns
documented by \citet{TodaWalsh2015JPE}, \citet{Lee2025}, and
\citet{GHWW2023}. The closed-form coefficient $\kappa(z)$ is also useful computationally: it provides a theoretically justified rule for extrapolating consumption beyond a finite wealth grid, as in the Pareto-extrapolation
approach of \citet{Gouin-BonenfantToda2023QE}.
\end{rem}

\subsection{Asymptotic MPCs for \texorpdfstring{$\delta \ge \gamma$}{}}\label{ss:pos_ampc}

We next study the cases $\delta\ge \gamma$ in the region $r(K(1-\gamma))<1$, where the stochastic mechanism in Proposition~\ref{prop:MPC0} is absent. In this region, the asymptotic MPC is strictly positive, and the role of wealth preferences depends on whether $\delta=\gamma$ or $\delta>\gamma$.

Recall $K(\theta)$ defined in \eqref{eq:Kmat}. For a given set $\AA$, let $\AA^\ZZ$ be the set of functions $f:\ZZ \to \AA$. For each $x\in  [1,\infty]^\ZZ$, define
\begin{equation}\label{eq:F}
    (Fx)(z) \coloneqq \begin{cases*}
    \infty & if $\delta < \gamma$, \\
    \left(1 + [K(1-\gamma)(x+\psi 1)](z)^{1/\gamma}\right)^\gamma  & if $\delta = \gamma$, \\
    \left(1 + [K(1-\gamma)x] (z)^{1/\gamma}\right)^{\gamma} & if $\delta > \gamma$,
    \end{cases*}
\end{equation}
where $(K(1-\gamma)x)(z) \coloneqq \E_z \hat{\beta} \hat{R}^{1-\gamma}x(\hat{Z})$ by definition.\footnote{Alternatively, in traditional matrix notation, $K(1-\gamma)x$ represents the column vector resulting from multiplying the matrix $K(1-\gamma)$ by the column vector $x$.}

\subsubsection{Asymptotic MPCs for \texorpdfstring{$\delta>\gamma$}{}}

To obtain a sharp characterization of this regime, we make the following regularity assumption.

\begin{asmp}[Return and income floor]\label{asmp:RYmin}
There exist $m_1,m_2>0$ such that, for all $(z,\hat{z})\in \ZZ^2$, we have $R(z,\hat{z},\hat{\epsilon})\in \set{0}\cup [m_1,\infty)$ and $Y(z,\hat{z},\hat{\epsilon})\ge m_2$ almost surely conditional on $(z,\hat{z})$.
\end{asmp}

The next theorem characterizes the asymptotic MPC when wealth utility is more concave than consumption utility.

\begin{thm}[Asymptotic irrelevance of wealth preferences]\label{thm:MPC2}
Suppose Assumptions~\ref{asmp:U}--\ref{asmp:RYmin} hold. If $\delta>\gamma$ and $r(K(1-\gamma))<1$, then $F$ in \eqref{eq:F} has a unique fixed point $x_2^*=(x_2^*(z))_{z\in \ZZ}\in [1,\infty)^\ZZ$, and for all $z\in \ZZ$, we have
\begin{equation}
    \lim_{w\to\infty} \frac{c^*(w,z)}{w}=x_2^*(z)^{-1/\gamma}. \label{eq:MPC2}
\end{equation}
\end{thm}

Theorem~\ref{thm:MPC2} shows that when $\delta>\gamma$, wealth preferences may affect consumption and saving decisions at finite wealth levels, but do not influence the asymptotic consumption-wealth ratio. Consequently, the asymptotic MPC and saving rate coincide with those in the benchmark model without wealth preferences.

\subsubsection{Asymptotic MPCs for \texorpdfstring{$\delta = \gamma$}{}}

Finally, we consider the knife-edge case in which consumption and wealth utilities exhibit identical curvature.

\begin{thm}[Asymptotic MPC under equal curvature]\label{thm:MPC1}
Suppose Assumptions~\ref{asmp:U}--\ref{asmp:Euler} hold with $u(c)=p(c;\gamma)$ and $v(w)=\psi p(w;\gamma)$, where $\gamma>0$ and $\psi\ge 0$. If $r(K(1-\gamma))<1$, then $F$ in \eqref{eq:F} has a unique fixed point $x_1^*=(x_1^*(z))_{z\in \ZZ}\in [1,\infty)^\ZZ$, and for all $z\in \ZZ$, we have
\begin{equation}
    \lim_{w\to\infty} \frac{c^*(w,z)}{w}=x_1^*(z)^{-1/\gamma}. \label{eq:MPC1}
\end{equation}
\end{thm}

\begin{rem}
When $\psi=0$, wealth does not enter utility and the fixed-point equation in Theorem~\ref{thm:MPC1} reduces to that in \citet{MaToda2021JET}. Hence, Theorem~\ref{thm:MPC1} recovers their positive asymptotic MPC result as a special case. Together with Proposition~\ref{prop:MPC0} (extended to permit $\psi=0$), our results recover their full characterization.
\end{rem}

Theorem~\ref{thm:MPC1} shows that when $\delta = \gamma$, wealth preferences remain relevant in the upper tail. In this case, the wealth-preference weight $\psi$ continues to affect the limiting consumption-wealth ratio through the fixed point $x_1^*$. However, unlike the case $\delta<\gamma$, wealth preferences do not generate a vanishing MPC or a different asymptotic scaling of consumption.

\subsection{Taxonomy of long-run consumption behavior}

Combining Theorems~\ref{thm:MPC0}--\ref{thm:MPC1} with Proposition~\ref{prop:MPC0} provides a complete characterization of the asymptotic consumption behavior under wealth preferences. The key implication of our theory arises when $\delta<\gamma$: wealth preferences generate vanishing asymptotic MPCs under the mild conditions of Theorem~\ref{thm:MPC0}, without requiring any restriction on the spectral radius of $K(1-\gamma)$. When $\delta\geq\gamma$, vanishing asymptotic MPCs may still arise through the spectral-radius mechanism characterized in Proposition~\ref{prop:MPC0}, which requires the restrictive condition $r(K(1-\gamma))\geq1$. In the complementary region $r(K(1-\gamma))<1$, the asymptotic MPC is governed by the relative curvature of wealth and consumption utility:
\begin{equation*}
    \lim_{w\to\infty}\frac{c^*(w,z)}{w}
    =
    \begin{cases*}
        0 & if $\delta<\gamma$,\\
        x_1^*(z)^{-1/\gamma} & if $\delta=\gamma$,\\
        x_2^*(z)^{-1/\gamma} & if $\delta>\gamma$,
    \end{cases*}
\end{equation*}
where the last case additionally requires
Assumption~\ref{asmp:RYmin}. The monotonicity of $F$ in \eqref{eq:F} implies $x_1^*\ge x_2^*$, so the asymptotic MPC under equal curvature is weakly lower than in the benchmark without wealth preferences. The weight $\psi$ determines the coefficient $\kappa(z)$ in \eqref{eq:consumption_power_law} when $\delta<\gamma$, affects the limiting MPC through $x_1^*$ when $\delta=\gamma$, and has no effect on the asymptotic MPC when
$\delta>\gamma$.

\begin{exmp}
Suppose $\beta_t\equiv\beta$ and $R_t\equiv R$. Then $F$ in
\eqref{eq:F} reduces to
\begin{equation*}
    Fx=
    \begin{cases*}
        \infty & if $\delta<\gamma$,\\
        \left(1+[\beta R^{1-\gamma}(x+\psi)]^{1/\gamma}\right)^\gamma
            & if $\delta=\gamma$,\\
        \left(1+[\beta R^{1-\gamma}x]^{1/\gamma}\right)^\gamma
            & if $\delta>\gamma$.
    \end{cases*}
\end{equation*}
Without wealth preferences, the asymptotic MPC is
\begin{equation*}
    \bar c_0=
    \begin{cases*}
        1-(\beta R^{1-\gamma})^{1/\gamma}
            & if $\beta R^{1-\gamma}<1$,\\
        0 & otherwise.
    \end{cases*}
\end{equation*}
Thus, under the standard impatience condition $\beta R<1$, a zero
asymptotic MPC in the benchmark model requires $R<1$. With wealth
preferences, the asymptotic MPC becomes
\begin{equation*}
    \bar c=
    \begin{cases*}
        0
            & if $\delta<\gamma$ or $\beta R^{1-\gamma}\ge 1$,\\
        (x_1^*)^{-1/\gamma}
            & if $\delta=\gamma$ and $\beta R^{1-\gamma}<1$,\\
        \bar c_0
            & if $\delta>\gamma$ and $\beta R^{1-\gamma}<1$,
    \end{cases*}
\end{equation*}
where $x_1^*$ is the unique fixed point of $F$ in $[1,\infty)$.
In particular, if $R\ge1$ and $\beta R<1$, then
Theorem~\ref{thm:MPC0} gives
\begin{equation*}
    \lim_{w\to\infty}
    \frac{c^*(w,z)}{w^{\delta/\gamma}}
    =
    \left(
        \frac{\beta R^{1-\delta}\psi}
             {1-\beta R^{1-\delta}}
    \right)^{-1/\gamma}.
\end{equation*}
Hence, unlike the benchmark stochastic mechanism, the wealth-preference mechanism generates a zero asymptotic MPC under
conventional positive-return and impatience conditions whenever
$\delta<\gamma$.
\end{exmp}

\subsection{Warm-glow bequests under random mortality}\label{subsec:bequest}

Our framework also admits an exact joy-of-giving/warm-glow bequest interpretation under random mortality in a perpetual-youth setting. Models combining mortality risk with direct utility from bequests date back at least to \citet{Hurd1989}. This provides a direct connection to the OLG models commonly used in quantitative work \citep{DeNardi2004,straub2019consumption,fella2024saving,Hubmer2026}, without requiring a separate finite-horizon analysis.

Let $q \in(0,1)$ be the probability of surviving from one period to the next, independently of other shocks, and let $\tau\in\N$ denote the date of death. Thus,
$\Pr(\tau>t)=q ^t$ for $t\ge 0$ and $\Pr(\tau=t)=(1-q)q ^{t-1}$ for $t\ge 1$. Let $\tilde\beta_t$ denote the discount factor before mortality
adjustment. We adopt the timing convention that, if death occurs at date $t$, wealth $w_t$ has already been realized but consumption $c_t$ has not been chosen, and the estate generates warm-glow utility $b(w_t)$. Consequently, after integrating out the independent mortality shock, the expected utility generated by a feasible policy $c$ through date $T$ is
\begin{equation}
    B_{c,T}(w,z)\coloneqq \E_{w,z}\left[\sum_{t=0}^T q ^t\left(\prod_{i=1}^t\tilde\beta_i\right)u(c_t)+\sum_{t=1}^T(1-q)q ^{t-1}\left(\prod_{i=1}^t \tilde\beta_i\right)b(w_t)
 \right]. \label{eq:bequest_utility}
\end{equation}

Define the mortality-adjusted discount factor and the effective utility from
wealth by $\beta_t\coloneqq q \tilde\beta_t$ and $v(w)\coloneqq \frac{1-q }{q }b(w)$. Then \eqref{eq:bequest_utility} becomes
\begin{align}
    B_{c,T}(w,z)&=\E_{w,z}\left[u(c_0)+\sum_{t=1}^T
     \left(\prod_{i=1}^t\beta_i\right)
     \left(u(c_t)+v(w_t)\right)\right] \notag\\
     &=V_{c,T}(w,z)-v(w), \label{eq:bequest_equivalence}
\end{align}
where $V_{c,T}$ is defined in \eqref{eq:Vc}.  Since the initial wealth $w$ is given, the last term in \eqref{eq:bequest_equivalence} is independent of the policy. Therefore, for any two feasible policies $c_1,c_2$ and every $T$,
\begin{equation*}
    B_{c_2,T}(w,z)-B_{c_1,T}(w,z)=V_{c_2,T}(w,z)-V_{c_1,T}(w,z).
\end{equation*}
Hence, the perpetual-youth bequest problem and the wealth-preference problem \eqref{eq:os} induce the same overtaking ordering and therefore the same optimal consumption policy. In effect, the survival probability is absorbed into the discount factor, while the death probability determines the effective weight on wealth utility.

In particular, if the primitive bequest utility is $b(w)=\psi_b p(w;\delta)$, then $v(w)=\psi p(w;\delta)$ with $\psi=\frac{1-q }{q }\psi_b$. Thus, after replacing $\tilde\beta_t$ by the mortality-adjusted discount factor $\beta_t=q \tilde\beta_t$, all optimality and asymptotic results in Theorems~\ref{thm:existence}--\ref{thm:MPC1} apply directly to this perpetual-youth bequest model. The comparison between $\delta$ and $\gamma$ has the same interpretation: when $\delta<\gamma$,
the warm-glow motive generates a vanishing asymptotic MPC and power-law
consumption with exponent $\delta/\gamma$; when $\delta=\gamma$, the bequest
motive affects the limiting MPC; and when $\delta>\gamma$, it is asymptotically irrelevant for the consumption-wealth ratio.

The exact equivalence above relies on geometric mortality and a stationary survival probability. In a deterministic finite-horizon model, age must be included in the state and the terminal bequest enters as a boundary condition. Although the terminal first-order condition displays the same comparison between $\delta$ and $\gamma$, continuation values affect consumption decisions at earlier stages, so the coefficient governing upper-tail consumption generally varies with age and is characterized by the finite-horizon recursion implied by Lemma~\ref{lem:Tc_bd}. The two formulations also differ in economic interpretation. A warm-glow bequest motive values the transfer of wealth at death, whereas direct utility from wealth can capture motives such as status or the enjoyment of holding wealth that do not require an intergenerational transfer. Consistent with this distinction, \citet{fella2024saving} find similar estimated bequest parameters for households with and without children, suggesting that persistent wealth holdings may reflect motives beyond purely altruistic transfers.

\section{Quantitative relevance}\label{sec:quant}

This section evaluates whether the wealth-preference mechanism characterized in Section~\ref{sec:asymptotic} is quantitatively relevant at empirically observed wealth levels. We use a parsimonious environment to assess three implications central to the theory: 
\begin{enumerate*}
    \item upper-tail wealth concentration relative to the classical benchmark,
    \item saving rates among wealthy households, and
    \item the accuracy of the asymptotic approximation.
\end{enumerate*}

\subsection{Model specification}

To isolate the wealth-preference mechanism, we use a partial equilibrium version of \citet[\S3]{BeareToda2025EI}, abstracting from portfolio choice and taxes but introducing wealth preferences and persistent-transitory income shocks. Unless stated otherwise, we use their parameter values.

\paragraph{Birth, death, and annuity}

We consider a perpetual-youth model as in Section~\ref{subsec:bequest}, which can be interpreted as either a wealth-preference model or a warm-glow bequest model. Let $q \in (0, 1)$ denote the probability of surviving to the next period, which we set to $q = 0.975$, implying an expected planning horizon (economic lifetime) of $1/(1-q)=40$ years. Deceased agents are replaced by newborn agents, who start with zero financial assets and hence wealth equal to income $w_t=Y_t>0$. There are perfectly competitive life insurance companies, so agents receive annuities while alive and surrender their wealth to the insurer upon death.

\paragraph{Income process}

Labor income $Y_t$ follows a discretized version of an exogenous stochastic process with a persistent component $Z_{1t}$ following an AR(1) process and a transitory component $\epsilon_{1t}$:
\begin{subequations}\label{eq:income}
\begin{align}
    Z_{1t} &= \lambda Z_{1t-1} + \nu_t, &  \set{\nu_t} &\iidsim N(0, \sigma_{\nu}^2), \label{eq:income1}\\
	\log Y_t &= Z_{1t} + \epsilon_{1t}, 
	& \set{\epsilon_{1t}} &\iidsim N(0,\sigma_1^2). \label{eq:income2}
\end{align}
\end{subequations}
One period corresponds to a year and we set $\lambda = 0.9735$, $\sigma_\nu = 0.1272$, and $\sigma_1 = 0.2102$ following \citet{lander2024estimating}. We discretize the persistent component into a 15-state Markov chain using the \citet{rouwenhorst1995} method, and the transitory component into a 7-point distribution using Gauss-Hermite quadrature.

\paragraph{Occupation and investment risk}

Occupation $Z_{2t}$ follows a two-state Markov chain independent of $Z_{1t}$ taking values in  $\set{\mathrm{w},\mathrm{e}}$, where w, e correspond to workers and entrepreneurs. The transition probabilities are $\Pr(\mathrm{w}\to \mathrm{e})=0.0026$ and $\Pr(\mathrm{e}\to \mathrm{w})=0.02$, which imply that 11.5\% of agents are entrepreneurs in the stationary distribution. A worker can invest only in a risk-free asset yielding a gross risk-free rate $R_f>0$. An entrepreneur can also invest in a risky asset yielding log excess return
\begin{equation}
    \log R_t-\log R_f\eqqcolon \epsilon_{2t}\iidsim N(\mu_2,\sigma_2^2). \label{eq:invest_shock}
\end{equation}
We set $\mu_2=0.035$ and $\sigma_2=0.247$ following Table 3 of \citet{FagerengGuisoMalacrinoPistaferri2020} and discretize the distribution of $\epsilon_{2t}$ into a 3-point distribution using Gauss-Hermite quadrature. We define the gross return on wealth of an entrepreneur by
\begin{equation*}
    R_f(1-\theta+\theta\exp(\epsilon_{2t})),
\end{equation*}
where $\theta$ is the portfolio share. We set $\theta=0.71$ from the private equity share of the top 0.1\% \citep[Table 1A]{FagerengGuisoMalacrinoPistaferri2020}. We collect the Markov state as $Z_t=(Z_{1t},Z_{2t})$ and the transitory shock as $\epsilon_t=(\epsilon_{1t},\epsilon_{2t})$. The state of a newborn agent is drawn from the stationary distribution.

\paragraph{Preference parameters}

Let $\beta$ and $R_f$ be the mortality-adjusted discount factor and gross risk-free rate: as in Section~\ref{subsec:bequest}, if $\tilde{\beta},\tilde{R}_f$ are the actual discount factor and gross risk-free rate, then $\beta=q \tilde{\beta}$ and $R_f=\tilde{R}_f/q $. We take $R_f$ as exogenous (partial equilibrium); this is without loss of generality because we may always reverse-engineer a production function in an \citet{Aiyagari1994} economy to achieve a particular $R_f$ in general equilibrium. We consider the power utility functions $u(c) = p(c;\gamma)$ and $v(w) = \psi p(w;\delta)$, where $p$ is as in \eqref{eq:CRRA} and $\psi>0$ is the strength of wealth preference. We set $\beta=0.96$ and $\gamma=1$ (log utility), which are standard. Logarithmic utility is not essential but is convenient because in the classical case of $\psi=0$ (no wealth preference), Example 2 of \citet{MaToda2021JET} implies that the asymptotic MPC is constant at $1-\beta$ regardless of other parameters.

We now calibrate $R_f$, $\delta$, and $\psi$. Assuming $\delta<\gamma$, by Theorem \ref{thm:MPC0}, the asymptotic law of motion for wealth is $w'=Rw$ (because consumption and income are negligible relative to wealth as $w\to\infty$). Let $X_{ij}$ be the log excess return conditional on transitioning from occupation $i$ to $j$ and define the matrix of conditional moment generating functions $\Psi(z)$ by $\Psi_{ij}(z)=\E[\exp(z X_{ij})]$. Let $P_2$ be the transition probability matrix of occupation. Letting $\zeta_w$ be the wealth Pareto exponent and applying the \citet{BeareToda2022ECMA} formula heuristically, $\zeta_w$ satisfies
\begin{equation}
    q R_f^{\zeta_w}r(P_2\odot \Psi(\zeta_w))=1, \label{eq:BeareToda}
\end{equation}
where $r$ denotes the spectral radius \eqref{eq:spectral_radius} and $\odot$ denotes the Hadamard (entrywise) product of matrices. We set the wealth Pareto exponent to $\zeta_w=1.4$ following \citet[Figure 3]{BeareToda2025EI} and solve \eqref{eq:BeareToda} for
\begin{equation*}
    R_f=(q r(P_2\odot \Psi(\zeta_w)))^{-1/\zeta_w}=0.9787.\footnote{A low return is natural here for two reasons. First, with wealth preferences and $\delta<\gamma$, the asymptotic wealth dynamics is $w'=Rw$. A finite-mean stationary wealth distribution therefore requires the mean effective return, including survival, to be below one. Second, $R_f$ is expressed in stationary units. In a growing economy, detrending divides the gross return by trend growth \citep[Equation (3.1b)]{MaToda2021JET}, lowering the return relative to its undetrended counterpart.}
\end{equation*}
By \eqref{eq:speed_conv}, the upper tail of the consumption distribution satisfies
\begin{equation*}
    \Pr(c>x)\sim \Pr(w^{\delta/\gamma}>x)=\Pr(w>x^{\gamma/\delta})\sim x^{-\zeta_w\gamma/\delta},
\end{equation*}
so consumption has Pareto exponent $\zeta_c=\zeta_w\gamma/\delta$. \citet{TodaWalsh2015JPE} estimate $\zeta_c=3.4$, so we set $\delta=(\zeta_w/\zeta_c)\gamma=0.4118$.

We calibrate $\psi$ by matching the wealth-earnings ratio at various percentiles. To this end, for each candidate parameter value $\psi$, we solve for the stationary distribution in the model.\footnote{We use the endogenous grid point method \citep{Carroll2006,MaToda2022JME} to compute the consumption function and Pareto extrapolation \citep{Gouin-BonenfantToda2023QE} to compute the stationary wealth distribution. The grid is exponential \citep{Gouin-BonenfantToda2023QE} with minimum grid point 0, maximum grid point $10^5$, median grid point 10, and number of grid points 100.} We then minimize the Euclidean distance between the model and empirical vectors of log wealth-earnings ratios, with the data taken from \citet[Table 1]{KuhnRios-Rull2025}, and obtain $\psi=\psihat$.\footnote{We note that the parameter $\psi$ is scale-dependent and hence has no natural interpretation unless the unit of wealth is defined. To see why, if we scale consumption and wealth by the factor $A>0$ and let $\psi_A$ be the parameter required to produce the same behavior, we need $\psi_A=\psi A^{\gamma-\delta}$.} Table~\ref{tab:parameters} summarizes the benchmark parameter values. Table~\ref{tab:wealth-earnings} compares the wealth-earnings ratios in the model against the empirical counterparts.

\begin{table}[!htb]
	\centering
	\caption{Benchmark parameter values.}
	\label{tab:parameters}
	\begin{tabular}{clc}
		\toprule
		Parameter & Description & Value \\
		\midrule
        $q $ & Survival probability in each period & 0.975 \\
		$\beta$ & Mortality-adjusted discount factor & 0.96 \\
		$\gamma$ & Relative risk aversion (consumption) & 1 \\
        $\delta$ & Relative risk aversion (wealth) & 0.4118 \\
		$\lambda$ & Autocorrelation of persistent income shock & 0.9735 \\
        $\sigma_\nu$ & Standard deviation of persistent income shock & 0.1272 \\
		$\sigma_1$ & Standard deviation of transitory income shock & 0.2102 \\
        $R_f$ & Mortality-adjusted gross risk-free rate & 0.9787 \\
        $\mu_2$ & Mean log excess return on risky asset & 0.035 \\
        $\sigma_2$ & Standard deviation of investment shock & 0.247 \\
        $\theta$ & Portfolio share of risky asset & 0.71 \\
        $\zeta_c$ & Pareto exponent (consumption) & 3.4 \\
        $\zeta_w$ & Pareto exponent (wealth) & 1.4 \\
        & $\Pr(\text{worker}\to \text{entrepreneur})$ & 0.0026 \\
        & $\Pr(\text{entrepreneur}\to \text{worker})$ & 0.02 \\
		$\psi$ & Strength of wealth preference & \psihat \\
		\bottomrule
	\end{tabular}
\end{table}

\begin{table}[!htb]
    \centering
    \caption{Empirical versus model-implied wealth-earnings ratios (targeted).}\label{tab:wealth-earnings}
    \begin{tabular}{lccccccc}
        \toprule
        Percentile & 40 & 50 & 60 & 80 & 90 & 95 & 99 \\
        \midrule
        Empirical & 3.40 & 3.97 & 4.52 & 6.91 & 9.42 & 12.45 & 14.04 \\
        Model & 3.85 & 4.31 & 4.75 & 6.51 & 7.74 & 10.81 & 15.08 \\
        \bottomrule
    \end{tabular}
\end{table}

\subsection{Wealth distribution}\label{ss:estim}

Table~\ref{tab:wealth_shares} evaluates an untargeted dimension of the model, the wealth distribution. Relative to the classical benchmark ($\psi=0$), the wealth-preference model ($\psi>0$) reduces the discrepancy from the data in every percentile group, although the model continues to understate the wealth share of the top 1\%.

\begin{table}[!htb]
	\centering
	\caption{Empirical versus model-implied wealth shares (non-targeted).}
	\label{tab:wealth_shares}
	\begin{tabular}{lcccccccc}
		\toprule
		Percentile & 0-20 & 20-40 & 40-60 & 60-80 & 80-90 & 90-95 & 95-99 & 99-100 \\
		\midrule
		Empirical & -0.002 & 0.010 & 0.038 & 0.101 & 0.120 & 0.124 & 0.259 & 0.351 \\
		Model ($\psi>0)$    & 0.019 & 0.050 & 0.095 & 0.177 & 0.154 & 0.118 & 0.163 & 0.224 \\
        Model ($\psi=0$) & 0.068 & 0.113 & 0.160 & 0.226 & 0.157 & 0.104 & 0.116 & 0.055 \\
		\bottomrule
	\end{tabular}
\end{table}

\begin{figure}[!htb]
    \centering
    \includegraphics[width=0.7\linewidth]{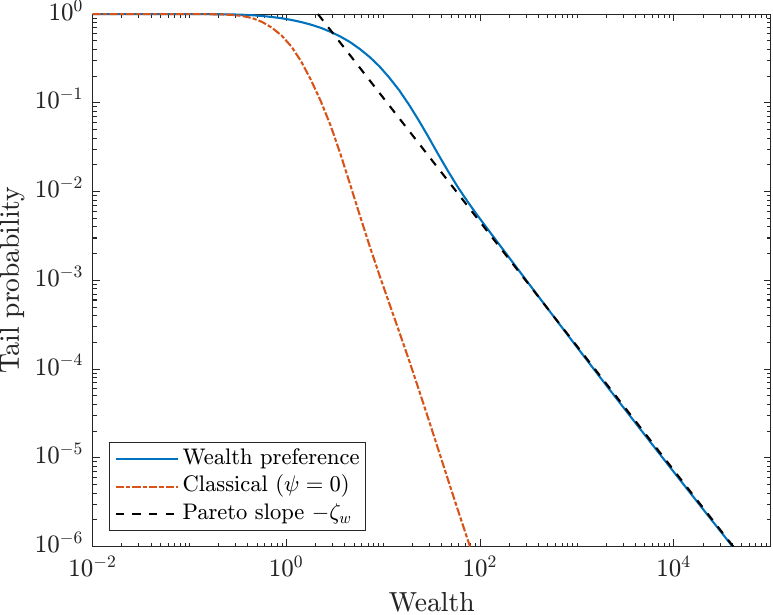}
    \caption{Wealth distribution.}
    \label{fig:wealth}
\end{figure}

Figure \ref{fig:wealth} shows the wealth distribution (top tail probability) on a log-log scale. By construction, the wealth-preference model ($\psi>0$) matches the wealth Pareto exponent $\zeta_w=1.4$. The classical model ($\psi=0$) also generates a Pareto tail due to idiosyncratic investment risk, but the Pareto exponent is much larger (2.9) and hence the top tail is much thinner.

\subsection{Consumption and saving}

We now assess the quantitative implications of the wealth-preference mechanism. While Section \ref{sec:asymptotic} characterizes the asymptotic behavior of consumption analytically, the relevance of these results depends on whether the asymptotic regime is reached at empirically meaningful wealth levels.

\begin{figure}[!htb]
	\includegraphics[width=\textwidth]{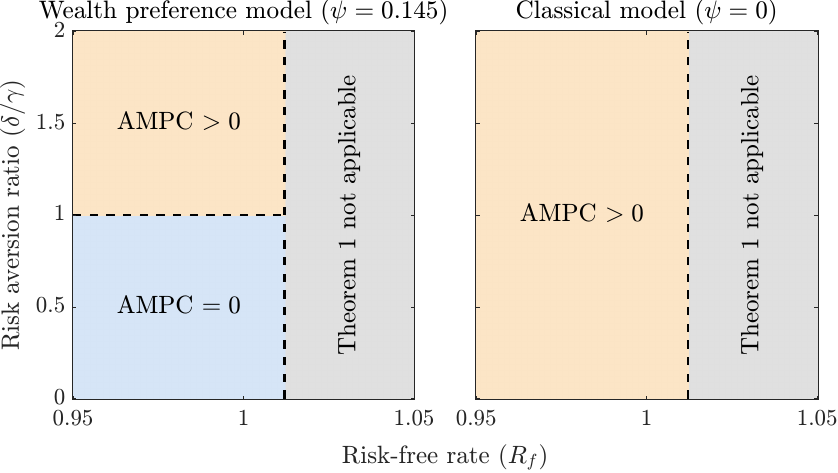}
	\caption{Phase diagram of asymptotic consumption regimes.}
	\label{fig:phase_diagram}
\end{figure}

We begin by characterizing the parameter regions associated with different asymptotic consumption regimes. Figure~\ref{fig:phase_diagram} illustrates these regions for the wealth-preference model ($\psi>0$) and the classical model ($\psi = 0$), with all other parameters held at their benchmark values reported in Table~\ref{tab:parameters}. In both cases, $r(K(1))<1$ holds (Assumption \ref{asmp:Euler}), so a unique optimal policy exists (Theorem \ref{thm:existence}). As we vary $\delta$, the asymptotic behavior of consumption undergoes a phase transition at $\delta = \gamma$ for the wealth-preference model. The shaded regions delineate parameter regions with zero versus positive asymptotic MPC (AMPC). The classical model displays a strikingly different pattern: the AMPC remains positive for all nonnegative net interest rates, highlighting that wealth preferences are an important source of persistent saving incentives among wealthy households.

\begin{figure}[!htb]
	\includegraphics[width=\textwidth]{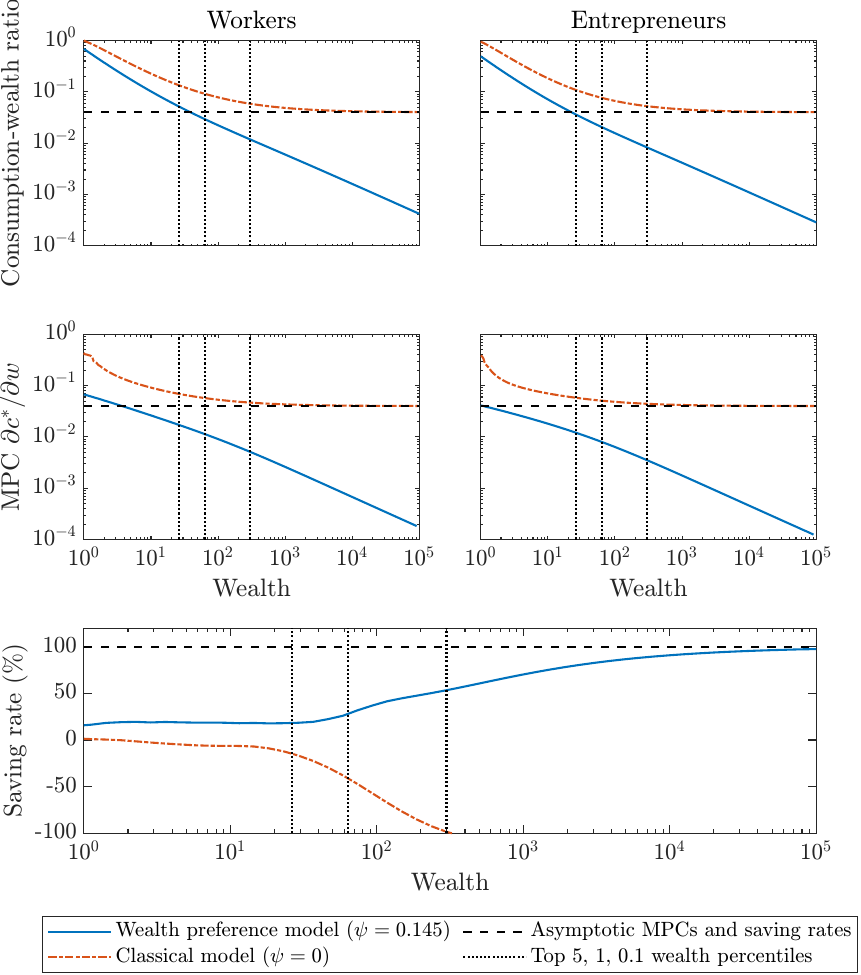}
	\caption{Consumption and saving rates (classical vs.\ wealth-preference models).}
	\label{fig:comp_psi}
\end{figure}

Next, we examine the MPC defined by $\partial c^*(w,z)/\partial w$ and saving rates across a broad wealth grid. The most natural definition of the saving rate is the fraction of wealth that remains after consumption, $1-c_t/w_t$, which is simply one minus the consumption-wealth ratio. Since wealth is difficult to observe, another common measure replaces the denominator with current income including capital gains/losses and aggregates across agents \citep{HuggettVentura2000,DynanSkinnerZeldes2004}, so
\begin{equation}
    s(w)\coloneqq 1-\frac{\E[c_{t+1}\mid w_t=w]}{\E[(R_{t+1}-1)(w_t-c_t)+Y_{t+1}\mid w_t=w]}. \label{eq:savingrate}
\end{equation}
Figure~\ref{fig:comp_psi} compares the wealth-preference model with the classical model. Here and elsewhere, we plot graphs corresponding to the median income state $(Z_{1t},\epsilon_{1t})=(0,0)$ in \eqref{eq:income2} (the results across income states are similar).

In the classical model, the MPC converges to the positive asymptotic constant $1-\beta$. Consequently, consumption continues to grow proportionally with wealth. In our calibration, this implies that the saving rate declines sharply with wealth and eventually becomes negative (below $-100$\%) at sufficiently high wealth levels. Such behavior contrasts with the empirical evidence that wealthy households maintain persistently high saving rates and accumulate wealth even at the upper tail of the distribution.

In contrast, the wealth-preference model generates markedly different upper-tail behavior. As illustrated in the top panel of Figure~\ref{fig:comp_psi}, the consumption-wealth ratio declines monotonically with wealth, consistent with our theoretical prediction of a vanishing MPC (Theorem \ref{thm:MPC0}). At the same time, the model generates persistently positive saving rates at high wealth levels, reflecting the continued incentive to accumulate wealth implied by the wealth-preference mechanism.

The vertical dotted lines indicate the top 5, 1, and 0.1 percentiles of the wealth distribution. In the wealth-preference model, the saving rates at these percentiles are 18\%, 28\%, and 53\%, respectively. The model reproduces the steep increase in saving rates across upper wealth percentiles, although the top 1\% level remains below the estimates of \citet{mian2025saving}. In this calibration, the asymptotic approximation is already accurate at these wealth levels.

\begin{figure}[!htb]
	\includegraphics[width=\textwidth]{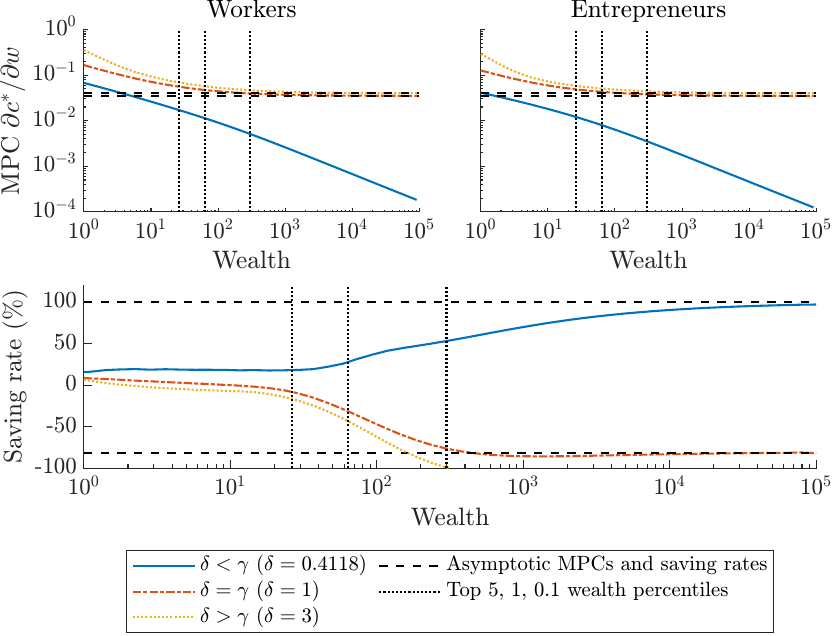}
	\caption{Consumption and saving rates (varying wealth risk aversion $\delta$).}
	\label{fig:comp_delta}
\end{figure}

Figure~\ref{fig:comp_delta} extends this analysis by varying $\delta$, the relative risk aversion with respect to wealth, holding the other parameters at their benchmark values reported in Table~\ref{tab:parameters}. Consistent with our theoretical results, when $\delta \ge \gamma$, the MPC converges to a positive limit, and saving rates become negative as wealth accumulates. Conversely, when $\delta < \gamma$, the AMPC vanishes, and the saving rate converges to $100\%$, consistent with the high saving rates of the rich.

\begin{figure}[!htb]
	\includegraphics[width=\textwidth]{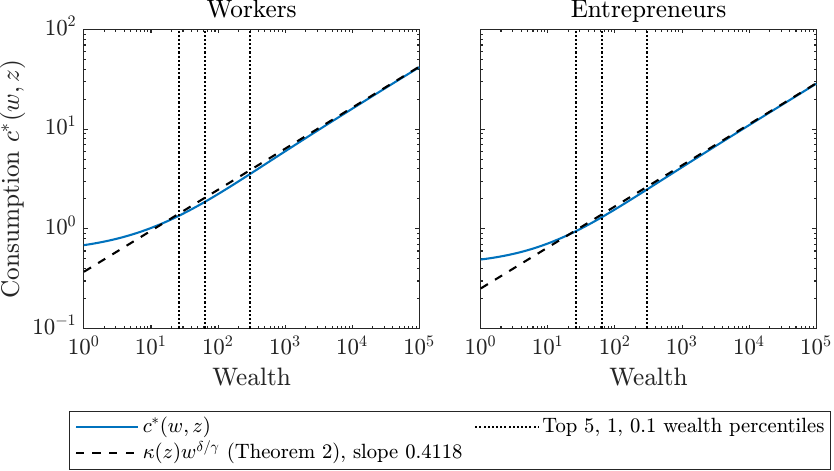}
	\caption{Asymptotic behavior of consumption ($\delta<\gamma$).}
	\label{fig:consumption_powerlaw}
\end{figure}

Finally, Theorem~\ref{thm:MPC0} establishes that $c^*(w, z) \approx \kappa(z) w^{\delta/\gamma}$ for $\delta < \gamma$, implying a power law between wealth and consumption. To quantify this, Figure~\ref{fig:consumption_powerlaw} plots the consumption function $c^*(w,z)$ and its asymptotic approximation against wealth on a log-log scale. Convergence to the asymptotic limit is rapid. By the time households reach the top 1 or 0.1 percentile of the wealth distribution, the consumption scaling is already close to the asymptotic prediction. These findings show that the asymptotic results derived in Section~\ref{sec:asymptotic} are not merely properties of the extreme tail at infinite wealth. Instead, they provide a quantitatively reliable approximation of consumption behavior over the range of wealth levels observed among the richest households.

\appendix

\section{Preliminaries}\label{ss:prelim}

Let $(\Omega, \fF, \Pr)$ be a fixed probability space on which all random variables are defined, $\set{\fF_t}_{t=0}^\infty\subset \fF$ be a filtration, and let $\E$ denote expectations under $\Pr$. Uncertainty is driven by two stochastic processes adapted to $\set{\fF_t}_{t=0}^\infty$. One is the time-homogeneous Markov chain $\set{Z_t}_{t=0}^\infty$ taking values in a finite set $\ZZ$, with transition probability $P(z,\hat{z})=\Pr[Z_t=\hat{z} \mid Z_{t-1}=z]$, and the other is the \iid innovation $\set{\epsilon_t}_{t=0}^\infty$ with marginal distribution $\pi$. Without loss of generality, we may assume $\ZZ = \set{1, \dots, Z}$. 

For $z\in \ZZ$, we write $\Pr_z$ for probability conditional on $Z_0 = z$ and $\E_z$ for expectation under $\Pr_z$. For any sets $A$ and $B$, we denote by $B^A$ the set of functions from $A$ to $B$. Following this convention, $\R^\ZZ$ and $\R_+^\ZZ$ denote the sets of real-valued and nonnegative real-valued functions on $\ZZ$, respectively. 

\subsection{Spectral radius inequality}

Recall that a linear operator $A : \R^\ZZ \to \R^\ZZ$ is called \emph{nonnegative} if it maps elements of $\R_+^\ZZ$ to itself. We first state a version of the Gelfand spectral radius formula, which connects the spectral radius of a linear operator to norms. A proof can be found in Theorem~9.1 of \citet{Krasnoselskii1972}.

\begin{lem}[Gelfand spectral radius formula]
	\label{lem:gelfand}
	Let $A:\R^\ZZ \to \R^\ZZ$ be a nonnegative linear operator, $h \in \R^\ZZ$ be a positive function, and $\norm{\cdot}$ be any norm on $\R^\ZZ$. Then the spectral radius of $A$ satisfies $r(A) = \lim_{n \to \infty} \norm{A^n h}^{1/n}$.
\end{lem}

For a nonnegative measurable function $\phi$, define the stochastic process $\phi_t = \phi(Z_{t-1},Z_t,\epsilon_t)$ and the associated matrix $L_\phi$ by
\begin{equation}\label{eq:lfunc}
    L_\phi(z,\hat{z}) \coloneqq P(z,\hat{z}) \int \phi(z,\hat{z},\hat{\epsilon}) \pi (\diff \hat{\epsilon}).
\end{equation}
The matrix $L_\phi$ is expressed as a function on $\ZZ \times \ZZ$ in \eqref{eq:lfunc}, but it can be represented in traditional matrix notation by enumerating $\ZZ$. In what follows, we treat $L_\phi$ as a linear operator on $\R^{\ZZ}$. A proof by induction shows that, for any $h \in \R^{\ZZ}$, 
\begin{equation}
    (L_\phi^n h)(z) = \E_z \left(\prod_{t=1}^n \phi_t \right) h(Z_n), \label{eq:ln_ope}
\end{equation}
where $L_\phi^n$ is the $n$-th composition of the operator $L_\phi$ with itself (equivalently, the $n$-th power of the matrix $L_\phi$). 

\begin{lem}[Spectral radius and long-run growth rates]
\label{lem:Lphi}
Let $\set{\phi_t}$ and $L_\phi$ be as defined in \eqref{eq:lfunc}. Then the following statements hold:
\begin{enumerate}
	\item\label{item:Lphi1} If $r(L_\phi) < 1$, then for any $\sigma\in (r(L_\phi),1)$, there exists $N \in \N$ such that $\max_{z \in \ZZ} \E_z \prod_{t=1}^n \phi_t < \sigma^n$ for all $n \ge N$.
	\item\label{item:Lphi2} If $\set{Z_t}$ is irreducible, then, letting $\E$ represent expectation under its unique stationary distribution, we have 
	$r(L_\phi) = \lim_{n\to\infty} \left(
	    \E \prod_{t=1}^{n} \phi_t
	\right)^{1/n}$.
\end{enumerate}
\end{lem}

\begin{proof}
By the nonnegativity of $L_\phi$ and Lemma~\ref{lem:gelfand}, for any norm $\norm{\cdot}$ on $\R^{\ZZ}$ and any strictly positive function $h\in \R_+^\ZZ$, we have
\begin{equation*}
	r(L_\phi) = \lim_{n \to \infty} \norm{L_\phi^n h}^{1/n}.
\end{equation*}
\medskip
\noindent \ref{item:Lphi1} Choosing $h\equiv 1$ and $\norm{f} = \max_{z\in \ZZ} \abs{f(z)}$, we obtain
\begin{equation*}
    r(L_\phi) = \lim_{n \to \infty} \left( \max_{z \in \ZZ} L_\phi^n 1(z) \right)^{1/n} = \lim_{n \to \infty} \left(\max_{z \in \ZZ} \E_z \prod_{t=1}^n \phi_t \right)^{1/n}.
\end{equation*}
Since $r(L_\phi)<1$, the claim follows directly from the definition of the limit.

\medskip
\noindent \ref{item:Lphi2} Letting $h\equiv 1$ and $\norm{f} \coloneqq \E\abs{f(Z_0)}$ gives
\begin{equation*}
	r(L_\phi) = \lim_{n \to \infty} \left(
	    \E \abs{\E_{Z_0} \prod_{t=1}^{n} \phi_t} 
	\right)^{1/n}
	= \lim_{n \to \infty} \left(
	    \E \prod_{t=1}^{n} \phi_t 
	\right)^{1/n}. \qedhere
\end{equation*}
\end{proof}

We will repeatedly use the following affine fixed-point lemma.

\begin{lem}[Fixed point characterization for an affine map]\label{lem:G_oper}
Let $A$ be a $\ZZ\times \ZZ$ nonnegative matrix with spectral radius $r(A)<1$ and $b\in \R_+^\ZZ$. Let $G:\R_+^\ZZ\to \R_+^\ZZ$ be the operator defined by $Gv=Av+b$. Then $G$ has a unique fixed point $v^*=(I-A)^{-1}b$ and $G^nv\to v^*$ for any $v\in \R_+^\ZZ$.
\end{lem}

\begin{proof}
Since $r(A)<1$, the Neumann series converges and yields $0\le \sum_{k=0}^\infty A^k=(I-A)^{-1}$. Define $v^*\coloneqq (I-A)^{-1}b$. Then $(I-A)v^*=b$, implying $Gv^*=Av^*+b=v^*$. Any fixed point of $G$ satisfies $Av+b=v$, so $v=(I-A)^{-1}b=v^*$. Finally, by Lemma~\ref{lem:gelfand}, we have $G^nv-v^*=A^n(v-v^*)\to 0$ since $r(A)<1$.
\end{proof}

\subsection{Perov contraction principle}
\label{sss:perov}
Let $\XX$ be a nonempty set, and let $\hH$ be a subset of all vector-valued functions $h: \XX \to \R^\ZZ$. A mapping $d:\hH\times \hH \to \R_+^\ZZ$ is called a \emph{vector-valued metric} if for all $f,g,h\in \hH$, we have:
\begin{enumerate*}
    \item $d(h,g) \ge 0$; 
    \item $d(h,g) = 0$ if and only if $h=g$;
    \item $d(h,g) = d(g,h)$; and
    \item $d(h,g) \le d(h,f) + d(f,g)$.
\end{enumerate*}
The pair $(\hH, d)$ is then called a \emph{vector-valued metric space}.

Consider the vector-valued metric $d = (d_1, \dots, d_Z)$ defined entrywise by
\begin{equation*}
	d_z(h, g) \coloneqq\sup_{x\in \XX} \abs{h_z(x) - g_z(x)}, 
	\quad z\in \ZZ, 
\end{equation*}
for all $h = (h_1, \dots, h_Z) \in \hH$ and $g=(g_1, \dots, g_Z) \in \hH$. Furthermore, define a scalar metric $\tilde d: \hH \times\hH \to \R_+$ by $\tilde d(h,g) \coloneqq\max_{z\in \ZZ} d_z(h,g)$ for all $h,g\in \hH$. Then $(\hH,\tilde d)$ is a standard metric space. We say that $(\hH, d)$ is \emph{complete} if the metric space $(\hH, \tilde d)$ is complete. In addition, we pair $\hH$ with the usual pointwise partial order, writing $h\le g$ if $h_z (x) \le g_z(x)$ for all $(x,z)\in \XX \times \ZZ$.

The following lemma is an extension of Blackwell's approach to contraction.

\begin{lem}[Perov contraction principle]\label{lem:Perov-Blackwell}
Let $(\hH, d)$ be a complete vector-valued metric space, and let $\tilde T:\hH\to\hH$ be a self-map satisfying the following properties:
\begin{enumerate}
	\item (Monotonicity) If $h\le g$, then $\tilde T h \le \tilde Tg$.
	\item (Discounting) There exists a nonnegative matrix $B$ with $r(B)<1$ such that, for all $h \in \hH$ and $a \in \R_+^\ZZ$, we have $h+a \in \hH$ and $\tilde T(h+a) \le \tilde T h + Ba$.
\end{enumerate}
Then the following statements hold:
\begin{enumerate}
	\item $d(\tilde T h, \tilde T g) \le B d(h,g)$ for all $h,g\in \hH$.
	\item $\tilde T$ has a unique fixed point $h^*$ in $\hH$.
	\item $\tilde d(\tilde T^n h, h^*) \to 0$ as $n\to \infty$ for all $h \in \hH$.
\end{enumerate}
\end{lem}

\begin{proof}
See Theorems~2--3 of \citet{Toda2021ORL}.
\end{proof}

\subsection{Fixed point theorem for monotone operators}

We pair $\R^\ZZ$ with the standard pointwise partial order. For $x,y\in \R^\ZZ$, we write $x \le (\ll)\ y$ if $x(z) \le (<)\ y(z)$ for all $z\in \ZZ$. 

Let $F:\R^\ZZ \to \R^\ZZ$ be an operator. We say that $F$ is \emph{monotone} if $Fx \le Fy$ for all $x,y\in \R^\ZZ$ with $x \le y$. Moreover, we say that $F$ is \emph{concave} if for all $\alpha\in(0,1)$ and $x,y\in \R^\ZZ$, $F(\alpha x + (1-\alpha) y) \ge \alpha Fx + (1-\alpha) Fy$, and $F$ is \emph{convex} if $-F$ is concave.

\begin{lem}[Fixed point theorem for monotone concave/convex operators]
\label{lem:mono_fpt}
Let $[v_0,v_1]$ be an order interval in $\R^\ZZ$ with $v_0\ll v_1$, and let $F: [v_0,v_1] \to [v_0,v_1]$ be a monotone operator. Suppose one of the following conditions holds:
\begin{enumerate}
	\item $F$ is a concave operator, $v_0 \ll Fv_0$, and $Fv_1 \le v_1$.
	\item $F$ is a convex operator, $v_0 \le Fv_0$, and $Fv_1 \ll v_1$.
\end{enumerate}
Then $F$ has a unique fixed point $x^* \in [v_0,v_1]$, and $F^n x \to x^*$ as $n\to \infty$ for all $x \in [v_0,v_1]$. 
\end{lem}

\begin{proof}
    See Theorem~2.1.2 of \citet{Zhang2013}.
\end{proof}

An application of Lemma~\ref{lem:mono_fpt} yields the following lemma.

\begin{lem}\label{lem:F_iter}
Let $A$ be a $\ZZ\times \ZZ$ nonnegative matrix and $b\in \R_+^\ZZ$. Define $\phi: \R_+ \to \R_+$ by $\phi(t)=(1+t^{1/\gamma})^\gamma$ and $F:\R_+^\ZZ\to \R_+^\ZZ$ by $Fx=\phi(Ax+b)$, where $\phi$ is applied entrywise. Then the following statements hold.
\begin{enumerate}
    \item\label{item:F_iter1} If $r(A)<1$, then $F$ has a unique fixed point $x^*$ in $\R_+^\ZZ$ and $F^nx\to x^*$ for any $x\in \R_+^\ZZ$.
    \item\label{item:F_iter2} If $A$ is irreducible and $r(A)\ge 1$, then $F$ has no fixed point in $\R_+^\ZZ$ and $F^nx\to \infty$ entrywise for any $x\in \R_+^\ZZ$. 
\end{enumerate}
\end{lem}

\begin{proof}
\ref{item:F_iter1} Since $\phi(t)/t\to 1$ as $t\to\infty$, for any $p>1$, we can take $q>0$ such that $\phi(t)<pt+q$ for all $t\ge 0$. Note that $q>0$ can be made arbitrarily large. Since $A$ is nonnegative and $r(A)<1$, take $p>1$ such that $pr(A)<1$ and corresponding $q>0$. Define $v_0,v_1\in \R_+^\ZZ$ by $v_0\coloneqq 0$ and
\begin{equation*}
    v_1\coloneqq (I-pA)^{-1}(pb+q1)\ge q1.
\end{equation*}
Clearly, $Fv_0=F0=\phi(b)\ge \phi(0)=1\gg 0=v_0$ and
\begin{equation*}
    Fv_1=\phi(Av_1+b)\ll p(Av_1+b)+q1=v_1.
\end{equation*}
Clearly $F$ is a monotone map. We can show that $\phi$ is convex when $\gamma \le 1$ and concave when $\gamma \ge 1$. Hence, $F$ is convex if $\gamma \le 1$ and concave if $\gamma \ge 1$.	By Lemma~\ref{lem:mono_fpt}, $F$ has a unique fixed point in the order interval $[v_0,v_1]$, and for any $x\in [v_0,v_1]$, the sequence $\set{F^nx}$ converges to this fixed point. To show the uniqueness of the fixed point of $F$ in $\R_+^\ZZ$, suppose $x_1^*,x_2^*$ are fixed points. Choose $q>0$ large enough that $x_1^*,x_2^*\in [v_0,v_1]$. Uniqueness of the fixed point on this interval gives $x_1^*=x_2^*$ so the fixed point is unique. Given initial $x\in \R_+^\ZZ$, choose $q>0$ large enough such that $x\in [v_0,v_1]$. Then $F^nx\to x^*$.

\medskip
\noindent \ref{item:F_iter2} We first show that $F$ has no fixed point in $\R_+^{\ZZ}$. Suppose to the contrary that $F$ has a fixed point $x^*\in \R_+^\ZZ$. Since $\phi > 0$, we have $x^* \gg 0$. Since clearly $\phi(t) > t$ for all $t \ge 0$, we have $x^* = Fx^* \ge \phi (Ax^*) \gg Ax^*$. Since $A$ is nonnegative irreducible, by the Perron-Frobenius theorem, we can take a left eigenvector $y\gg 0$ such that $y'A = r(A) y'$. Since $x^* \gg A x^*$ and $y \gg 0$, we obtain $r(A) y' x^* = y' A x^* < y' x^*$. Dividing both sides by $y'x^* > 0$, we obtain $r(A) < 1$, which is a contradiction. Hence, $F$ has no fixed point in $\R_+^\ZZ$.

We next show that $F^nx\to \infty$ entrywise for any $x\in \R_+^\ZZ$. Let $x_n\coloneqq F^nx$ and $y_n\coloneqq F^n0$. The monotonicity of $F$ implies $x_n\ge y_n$, so it suffices to show that $y_n(z)\to \infty$ for all $z\in \ZZ$. Since $y_1=F0\ge 1\gg 0=y_0$, the monotonicity of $F$ implies that $\set{y_n}$ is an increasing sequence, i.e., $y_n \ge y_{n-1}$ for all $n$. Since $\set{y_n}$ is increasing in $\R_+^\ZZ$, if it were bounded, then it would converge to some $x^* \in \R_+^{\ZZ}$. By continuity of $F$, the limit $x^*$ would satisfy $F x^* = x^*$, contradicting the absence of fixed points. Therefore, $\set{y_n}$ must be unbounded, so there exists $\hat z \in \ZZ$ such that $y_n(\hat z) \to \infty$.
	
Finally, we use the irreducibility of $A$ to propagate this divergence to all $z\in \ZZ$. For any $(z,\hat z) \in \ZZ^2$, irreducibility implies that there exists $m \in \N$ such that $A^m_{z \hat z} > 0$. Hence, as $n \to \infty$, $y_{m+n} (z) \ge A_{z\hat{z}}^m y_n(\hat{z}) \to \infty$. It follows that $y_n(z) \to \infty$ for all $z \in \ZZ$, completing the proof.
\end{proof}

Lemma~\ref{lem:F_iter} extends \citet[Proposition~14]{MaToda2021JET}, which corresponds to the case $b=0$. A related treatment of power-transformed affine systems is provided by \citet{StachurskiWilmsZhang2025}, whose results also cover the $b=0$ case under irreducibility. Our lemma allows a nonnegative term $b$ inside the power transformation, permits reducible $A$ in part~\ref{item:F_iter1}, and establishes entrywise divergence in part~\ref{item:F_iter2}.

\section{Proof of Section~\ref{sec:os} results}

\subsection{First-order and transversality conditions}\label{ss:foc_tvc}

To derive the Euler equation and the transversality condition in Section~\ref{ss:opt_conds}, we apply optimal control theory \citep[Chapter~15]{TodaEME}. Consider the abstract optimal control problem
\begin{align*}
    &\maximize && \E_0\sum_{t=0}^\infty r_t(x_t,a_t)\\
    &\st && x_{t+1}=g_{t+1}(a_t) \quad \text{and} \quad a_t\ge 0,
\end{align*}
where $x_0$ is given, $x_t$ and $a_t$ denote the state and control variables, $r_t$ is the reward function, and $g_t$ is the transition function. Suppose $r_t$ is concave with $D_xr_t\ge 0$ and $D_ar_t\le 0$, where $D_x$ denotes the Jacobian with respect to $x$, and that $g_t$ is concave. Then the \emph{first-order condition} for optimality is
\begin{equation}
    D_ar_t(x_t,a_t)+\E_t\set{D_xr_{t+1}(x_{t+1},a_{t+1})D_ag_{t+1}(a_t)}\le 0 \label{eq:foc_abstract}
\end{equation}
with equality if $a_t>0$. The associated \emph{transversality condition} is
\begin{equation}
    \lim_{t\to\infty}\E_0[D_ar_t(x_t,a_t)a_t]=0. \label{eq:tvc_abstract}
\end{equation}
Under the concavity assumptions above, the first-order condition together with the transversality condition is sufficient for optimality \citep[Theorem 15.3]{TodaEME}.

For the optimal saving problem, let the state variable be wealth $w_t$ and control variable be saving $s_t=w_t-c_t$, so that $(x_t,a_t)=(w_t,s_t)$. The reward and transition functions are given by
\begin{equation*}
    r_t(w_t,s_t)=\left(\prod_{i=1}^t \beta_i\right)[u(w_t-s_t)+v(w_t)]
    \quad \text{and} \quad 
    g_{t+1}(s_t)=R_{t+1}s_t+Y_{t+1}.
\end{equation*}
Applying \eqref{eq:foc_abstract}, dividing by $\prod_{i=1}^t\beta_i$ (the case $\beta_i=0$ causes no problem because we may consider conditional Euler equations), and substituting $(c_t,w_t)=(c(w,z),w)$ and $\hat{w} = \hat{R} \left(w - c(w,z) \right) + \hat{Y}$, we obtain the Euler inequality
\begin{equation*}
\left( u'\circ c \right)(w,z) \ge 
    \E_z \hat{\beta} \hat{R} (u' \circ c + v') (\hat{w}, \hat{Z})
\end{equation*}
for all $(w,z) \in \SS$, and equality holds whenever $c(w,z) < w$. Since $u'$ is decreasing, this condition can be equivalently written in the compact form given by the Euler equation \eqref{eq:foc}. Finally, substituting the optimal saving problem into \eqref{eq:tvc_abstract} yields the transversality condition \eqref{eq:tvc}.

\subsection{Proof of Theorem \ref{thm:existence}}

We prove Theorem~\ref{thm:existence} by establishing a series of lemmas. In what follows, Assumptions~\ref{asmp:U} and \ref{asmp:Euler} are in force.

\begin{lem}\label{lem:metric}
$(\cC, \rho)$ is a complete metric space.
\end{lem}

\begin{proof}
The proof is identical to that of Proposition~4.1 of \citet{LiStachurski2014}, after replacing their asset variable by $w$, and is omitted.
\end{proof}

The following lemma shows that the time iteration operator $T$ is well defined.

\begin{lem}[Well-defined operator]
\label{lem:welldef_T}
For all $c \in \cC$ and $(w,z) \in \SS$, there exists a unique $\xi \in (0,w]$ that solves \eqref{eq:T_opr}.
\end{lem}

\begin{proof}
Fix $c \in \cC$ and $(w,z) \in \SS$. Since $c \in \cC$, the map $\xi \mapsto \psi_c(\xi, w, z)$ is increasing. Additionally, because $\xi \mapsto u'(\xi)$ is strictly decreasing by Assumption~\ref{asmp:U}, the equation \eqref{eq:T_opr} can have at most one solution. Therefore, uniqueness holds.

To establish existence, we apply the intermediate value theorem. It suffices to verify the following three conditions:
\begin{enumerate}[(a)]
  \item\label{item:welldef_a} the map $\xi \mapsto \psi_c(\xi, w, z)$ is continuous on $(0, w]$,
  \item\label{item:welldef_b} there exists $\xi \in (0,w]$ such that $u'(\xi) \ge \psi_c(\xi, w, z)$, and
  \item\label{item:welldef_c} there exists $\xi \in (0,w]$ such that $u'(\xi) \le \psi_c(\xi, w, z)$.
\end{enumerate}

\medskip
\noindent \ref{item:welldef_a} It suffices to show that $\xi \mapsto g_c(\xi,w,z)$ is continuous on $(0,w]$. To this end, fix $\xi \in (0,w]$ and let $\xi_n \to \xi$. By condition \eqref{eq:C4}, $c\in\cC$ implies the existence of a constant $M \in \R_+$ such that 
\begin{equation}
    u'(w) \le (u' \circ c)(w,z) \le u'(w) + M \quad \text{for all } (w,z) \in \SS. \label{eq:bd_uprime}
\end{equation}
By \eqref{eq:bd_uprime} and the monotonicity of $u'$ and $v'$, we have
\begin{subequations}\label{eq:uppbd_ruprmc}
\begin{align}
    \hat{\beta} \hat{R} \left( u' \circ c \right) 
        (\hat{R} \left(w - \xi \right) + \hat{Y}, \hat{Z})
    &\le \hat{\beta} \hat{R} \left( u' \circ c \right) ( \hat{Y}, \hat{Z} )
    \le \hat{\beta} \hat{R} u'( \hat{Y}) + \hat{\beta} \hat{R} M, \\
    \hat{\beta}\hat{R} v'(\hat{R}(w-\xi) + \hat{Y}) 
	&\le \hat{\beta} \hat{R} v'(\hat{Y}).
\end{align}
\end{subequations}
By Assumption~\ref{asmp:Euler}\ref{item:Euler1}, we know that $\E_z \hat{\beta} \hat{R} < \infty$, $\E_z \hat{\beta} \hat{R} u'(\hat{Y}) < \infty$, and $\E_z \hat{\beta} \hat{R} v'(\hat{Y}) < \infty$. Hence, by the dominated convergence theorem and the continuity of $c$, we 
conclude that $g_c(\xi_n,w,z) \to g_c(\xi,w,z)$. This proves that $\xi \mapsto 
\psi_c(\xi, w, z)$ is continuous.

\medskip
\noindent \ref{item:welldef_b} The claim clearly holds, since $u'(\xi) \to \infty$ as $\xi \to 0$ and $\xi \mapsto \psi_c(\xi, w, z)$ is increasing and always finite (since it is continuous as shown in the previous paragraph). 

\medskip
\noindent \ref{item:welldef_c} This follows by setting $\xi=w$.
\end{proof}

The following lemma shows that $T$ is a self-map on $\cC$.

\begin{lem}[Self-map]
\label{lem:self_map}
We have $Tc \in \cC$ for all $c \in \cC$.
\end{lem}

\begin{proof}
Fix $c \in \cC$ and let $g_c$ be defined as in
\eqref{eq:keypart_T_opr}. Set
$M\coloneqq\norm{u'\circ c-u'}<\infty$, so \eqref{eq:bd_uprime} holds,
and define
\begin{equation*}
    A_z \coloneqq \E_z \hat{\beta}\hat{R}
    \left(u'(\hat{Y})+v'(\hat{Y})+M\right),
    \qquad
    \bar A \coloneqq \max_{z\in\ZZ} A_z < \infty.
\end{equation*}
The finiteness of $\bar A$ follows from
Assumption~\ref{asmp:Euler}\ref{item:Euler1} and the finiteness of $\ZZ$.
It suffices to verify the following four conditions:
\begin{enumerate*}[(a)]
    \item\label{item:self_a} $Tc$ is continuous,
    \item\label{item:self_b} $Tc$ is increasing in $w$,
    \item\label{item:self_c} $0<Tc(w,z)\le w$, and
    \item\label{item:self_d} $\norm{u'\circ (Tc)-u'}<\infty$.
\end{enumerate*}

\medskip
\noindent \ref{item:self_a}
We first show that $g_c$ is jointly continuous on the set $G$ defined in
\eqref{eq:dom_T_opr}. Let $\set{(\xi_n,w_n,z_n)}\subset G$ satisfy
$(\xi_n,w_n,z_n)\to(\xi,w,z)\in G$. Since $\ZZ$ is finite and hence
discrete, $z_n=z$ for all sufficiently large $n$. Thus, without loss of
generality, suppose $z_n=z$ for all $n$. Define
\begin{equation*}
    \hat w_n \coloneqq \hat R(w_n-\xi_n)+\hat Y
    \quad\text{and}\quad
    \hat w \coloneqq \hat R(w-\xi)+\hat Y.
\end{equation*}
Then $\hat w_n\to\hat w$ almost surely. By the continuity of $c$, $u'$, and
$v'$,
\begin{equation*}
    \hat\beta\hat R (u'\circ c+v')(\hat w_n,\hat Z)
    \to
    \hat\beta\hat R (u'\circ c+v')(\hat w,\hat Z)
\end{equation*}
almost surely. Moreover, since $\hat w_n\ge\hat Y$, it follows from
\eqref{eq:uppbd_ruprmc} that
\begin{equation*}
    0\le
    \hat\beta\hat R (u'\circ c+v')(\hat w_n,\hat Z)
    \le
    \hat\beta\hat R
    \left(u'(\hat Y)+v'(\hat Y)+M\right).
\end{equation*}
The random variable on the right-hand side is integrable under $\Pr_z$ by
Assumption~\ref{asmp:Euler}\ref{item:Euler1}. Hence, the dominated convergence
theorem implies
\begin{equation*}
    g_c(\xi_n,w_n,z_n)\to g_c(\xi,w,z).
\end{equation*}
Therefore, $g_c$ is jointly continuous on $G$, and so is
$\psi_c(\xi,w,z)=\max\set{g_c(\xi,w,z),u'(w)}$.

We now prove that $Tc$ is continuous. Define
\begin{equation*}
    \Phi_c(\xi,w,z)\coloneqq u'(\xi)-\psi_c(\xi,w,z).
\end{equation*}
For each $(w,z)\in\SS$, the map $\xi\mapsto\Phi_c(\xi,w,z)$ is strictly
decreasing on $(0,w]$, and $Tc(w,z)$ is its unique zero by
Lemma~\ref{lem:welldef_T}. Let $(w_n,z_n)\to(w,z)$ in $\SS$, and set
$\xi_n\coloneqq Tc(w_n,z_n)$ and $\xi\coloneqq Tc(w,z)$. Again, we may
suppose that $z_n=z$ for all $n$.

Fix $\epsilon\in(0,\xi)$. Since $\xi-\epsilon<w$, the point
$(\xi-\epsilon,w_n,z)$ belongs to $G$ for all sufficiently large $n$. Since
$\Phi_c(\xi-\epsilon,w,z)>0$, joint continuity implies
$\Phi_c(\xi-\epsilon,w_n,z)>0$ for all sufficiently large $n$. Since $\Phi_c$ is strictly decreasing in its first argument, $\xi_n>\xi-\epsilon$. If $\xi<w$, take in addition
$\epsilon<w-\xi$. Then $(\xi+\epsilon,w_n,z)\in G$ for all sufficiently
large $n$. Since $\Phi_c(\xi+\epsilon,w,z)<0$, joint continuity implies
$\Phi_c(\xi+\epsilon,w_n,z)<0$ for all sufficiently large $n$, and hence
$\xi_n<\xi+\epsilon$. Therefore, $\xi_n\to\xi$ when $\xi<w$. If
$\xi=w$, the lower bound gives $\liminf_n\xi_n\ge w$, while
$\xi_n\le w_n\to w$ gives $\limsup_n\xi_n\le w$. Thus, $\xi_n\to\xi$ in
this case as well. Hence, $Tc$ is continuous on $\SS$.

\medskip
\noindent \ref{item:self_b}
Suppose that for some $z\in\ZZ$ and $w_1,w_2\in(0,\infty)$ with $w_1<w_2$,
we have
\begin{equation*}
    \xi_1\coloneqq Tc(w_1,z)>Tc(w_2,z)\eqqcolon\xi_2.
\end{equation*}
Since $c$ is increasing in wealth while $u'$ and $v'$ are decreasing,
$\psi_c$ is increasing in $\xi$ and decreasing in $w$. Therefore,
\begin{equation*}
    u'(\xi_1)<u'(\xi_2)
    =\psi_c(\xi_2,w_2,z)
    \le\psi_c(\xi_1,w_1,z)
    =u'(\xi_1),
\end{equation*}
which is a contradiction. Hence, $Tc$ is increasing in $w$.

\medskip
\noindent \ref{item:self_c}
Lemma~\ref{lem:welldef_T} implies that $Tc(w,z)\in(0,w]$ for all
$(w,z)\in\SS$.

\medskip
\noindent \ref{item:self_d}
Let $\xi\coloneqq Tc(w,z)$. Since $\xi\le w$ and $u'$ is decreasing,
\eqref{eq:T_opr} and \eqref{eq:uppbd_ruprmc} imply
\begin{align*}
    0 &\le u'(\xi)-u'(w) = \max\set{g_c(\xi,w,z),u'(w)}-u'(w) \\
    &\le g_c(\xi,w,z)
    \le A_z
    \le\bar A.
\end{align*}
Taking the supremum over $(w,z)\in\SS$ gives
$\norm{u'\circ(Tc)-u'}\le\bar A<\infty$.
\end{proof}

To prove Theorem~\ref{thm:existence}, we apply the Perov contraction principle (Lemma~\ref{lem:Perov-Blackwell}). To that end, it is more convenient to work with the space of marginal utilities instead of consumption functions. Let $\hH$ denote the set of all continuous functions $h:(0,\infty)\to \R^{\ZZ}$ such that each $h_z$ (where we write $h(w)\coloneqq (h_1(w),\dots,h_Z(w))$) is decreasing and $w\mapsto h_z(w)-u'(w)$ is bounded and nonnegative. For any $h \in \hH$, define $(\tilde{T} h)_z(w)$ as the value $\kappa>0$ that solves
\begin{equation}
\kappa = \max \set{ 
    \E_z \, \hat{\beta} \hat{R}
    (h+v') (\hat{R} [w - (u')^{-1}(\kappa)] + \hat{Y}, \hat{Z}), u'(w)}, \label{eq:kappa}
\end{equation}
where (with a slight abuse of notation) we define
\begin{equation*}
	(h+v')(w,z) \coloneqq h_z(w) + v'(w).
\end{equation*}
Moreover, consider the bijection $H: \cC \to \hH$ defined by $(Hc)_z(w) \coloneqq u'(c(w,z))$.

\begin{lem}[Conjugate operator]
\label{lem:conjug}
The operator $\tilde{T}$ is a self-map on $\hH$ and satisfies $\tilde{T}H = HT$ on $\cC$.
\end{lem}

\begin{proof}
Pick any $c \in \cC$ and $(w,z) \in \SS$. Let $\xi \coloneqq Tc(w,z)$. By definition (see \eqref{eq:T_opr}), $\xi$ solves
\begin{equation}
    u'(\xi) = 
     \max \set{\E_z \hat{\beta} \hat{R}(u' \circ c + v')(\hat{R} (w - \xi) + \hat{Y}, \hat{Z}), u'(w)}. \label{eq:Tc_eq}
\end{equation}
We need to show that $HTc$ and $\tilde{T} Hc$ evaluate to the same number at $(w,z)$. In other words, we need to verify that $u'(\xi)$ is the solution to
\begin{equation*}
    \kappa = \max \set{\E_z \hat{\beta} \hat{R}( u' \circ c + v')(\hat{R} [w - (u')^{-1} (\kappa)] + \hat{Y}, \hat{Z}), u'(w)}.
\end{equation*}
But this follows immediately from \eqref{eq:kappa} and \eqref{eq:Tc_eq}. Hence, we have shown that $\tilde{T} H = H T$ on $\cC$. Since $H \colon \cC \to \hH$ is a bijection,
we have $\tilde{T} = HT H^{-1}$. Moreover, Lemma~\ref{lem:self_map} 
ensures that $T \colon \cC \to \cC$, and hence $\tilde{T} \colon \hH \to \hH$. This completes the proof.
\end{proof}

To apply Lemma~\ref{lem:Perov-Blackwell}, for $h,g\in \hH$ and $z\in \ZZ$, we define
\begin{equation*}
    d_z(h,g)\coloneqq \sup_{w\in (0,\infty)}\abs{h_{z}(w)-g_{z}(w)}.
\end{equation*}
By the definition of $\hH$, $d_z$ is always finite. Moreover, define the vector-valued metric $d:\hH\times \hH\to \R_+^{\ZZ}$ by $d(h,g)=(d_1(h,g),\dots,d_Z(h,g))$. Let $\norm{\cdot}$ denote the supremum norm on $\R^{\ZZ}$.

\begin{lem}[Perov contraction]\label{lem:Perov-os}
Let $B\coloneqq K(1)$. Then we have:
\begin{enumerate}
	\item $d(\tilde{T}h_1,\tilde{T}h_2)\le Bd(h_1,h_2)$ for all $h_1,h_2\in \hH$.
	\item $\tilde T$ has a unique fixed point $h^*$ in $\hH$.
	\item For all $h\in \hH$, $\lVert d(\tilde T^n h, h^*)\rVert \to 0$ as $n\to \infty$.
\end{enumerate}
\end{lem}

\begin{proof}
By Assumption \ref{asmp:Euler}\ref{item:Euler1}, we have $r(B)\in [0,1)$. To show that the conclusions are true, by Lemma~\ref{lem:Perov-Blackwell}, it suffices to verify the following two conditions:
\begin{enumerate}[(a)]
    \item\label{item:Perov_a} If $h_1,h_2\in \hH$ and $h_1\le h_2$ pointwise, then $\tilde{T}h_1\le \tilde{T}h_2$ pointwise.
    \item\label{item:Perov_b} For any $h\in \hH$ and $a\in \R_+^{\ZZ}$, we have $\tilde{T}(h+a)\le \tilde{T}h+Ba$ pointwise.
\end{enumerate}

\medskip
\noindent \ref{item:Perov_a} Suppose $h_1\le h_2$ and take any $(w,z)\in \SS$. For $i\in\set{1,2}$, let $\kappa_i=(\tilde{T}h_i)_z(w)$, which satisfies \eqref{eq:kappa}. Let $\xi_i=(u')^{-1}(\kappa_i)$. Since $u'$ is strictly decreasing, to prove $\kappa_1\le \kappa_2$, it suffices to show $\xi_1\ge \xi_2$. Hence suppose to the contrary that $\xi_1<\xi_2$. By Lemma \ref{lem:conjug} and \eqref{eq:kappa}, we obtain
\begin{align*}
    u'(\xi_1)&=\max\set{\E_z \hat{\beta}\hat{R}(h_1+v')(\hat{R}(w-\xi_1)+\hat{Y},\hat{Z}),u'(w)}\\
    &\le \max\set{\E_z \hat{\beta}\hat{R}(h_2+v')(\hat{R}(w-\xi_1)+\hat{Y},\hat{Z}),u'(w)}\\
    &\le \max\set{\E_z \hat{\beta}\hat{R}(h_2+v')(\hat{R}(w-\xi_2)+\hat{Y},\hat{Z}),u'(w)}=u'(\xi_2),
\end{align*}
where the first inequality uses $h_1\le h_2$ and the second inequality uses the fact that $h_2,v'$ are decreasing in $w$ and $\xi_1<\xi_2$. Since $u'$ is strictly decreasing, we obtain $\xi_1\ge \xi_2$, which is a contradiction.

\medskip
\noindent \ref{item:Perov_b} Let $h\in \hH$ and $a\in \R_+^{\ZZ}$. Fixing $(w,z)\in \SS$, let $\kappa(a)$ be the value of $\kappa$ in \eqref{eq:kappa} where $h$ is replaced by $h+a$. Let $\xi(a)=(u')^{-1}(\kappa(a))$. Since $\tilde{T}$ is monotonic by part~\ref{item:Perov_a}, we have $\xi(a)\le \xi(0)$. Therefore,
\begin{align*}
    &(\tilde{T}(h+a))_z(w)=\kappa(a)=u'(\xi(a))\\
    &=\max\set{\E_z \hat{\beta}\hat{R}(h+a+v')(\hat{R}(w-\xi(a))+\hat{Y},\hat{Z}),u'(w)}\\
    &=\max\set{\E_z \hat{\beta}\hat{R}(h+v')(\hat{R}(w-\xi(a))+\hat{Y},\hat{Z})+\E_z\hat{\beta}\hat{R}a_{\hat{Z}},u'(w)}\\
    &\le \max\set{\E_z \hat{\beta}\hat{R}(h+v')(\hat{R}(w-\xi(a))+\hat{Y},\hat{Z}),u'(w)}+\E_z\hat{\beta}\hat{R}a_{\hat{Z}}\\
    &\le \max\set{\E_z \hat{\beta}\hat{R}(h+v')(\hat{R}(w-\xi(0))+\hat{Y},\hat{Z}),u'(w)}+\E_z\hat{\beta}\hat{R}a_{\hat{Z}}\\
    &=(\tilde{T}h)_z(w)+(Ba)_z.
\end{align*}
Therefore, $\tilde{T}(h+a)\le \tilde{T}h+Ba$ pointwise.
\end{proof}

\begin{proof}[Proof of Theorem \ref{thm:existence}]
\ref{item:existence1} Lemma~\ref{lem:self_map} implies that $T:\cC\to \cC$. If $c_1\le c_2$ pointwise, then the topological conjugacy result in Lemma~\ref{lem:conjug} and the monotonicity of $\tilde{T}$ in part \ref{item:Perov_a} of the proof of Lemma \ref{lem:Perov-os} imply that $Tc_1\le Tc_2$ pointwise, so $T$ is monotone. By Lemma~\ref{lem:Perov-os}, there exists a unique fixed point $h^*\in \hH$ of $\tilde{T}$. From the topological conjugacy result in Lemma~\ref{lem:conjug}, we have $\tilde{T} = H T H^{-1}$, so there exists a unique fixed point $c^*\in \cC$ of $T$.

\medskip
\noindent \ref{item:existence2} Take any $\sigma \in (r(B),1)$. With a slight abuse of notation, let $\norm{\cdot}$ denote the supremum norm on $\R^\ZZ$, the operator norm it induces on $\R^{\ZZ\times\ZZ}$, and the supremum norm on the space of functions from $(0,\infty)$ to $\R^\ZZ$. Since $\tilde{T}^n=HT^nH^{-1}$, it follows from the definition of $\rho$, $\cC$, $\hH$, and Lemmas~\ref{lem:Lphi} and \ref{lem:Perov-os} that
\begin{align*}
    \rho(T^n c, c^*)&=\rho(T^n c, T^n c^*)=\norm{HT^nc-HT^nc^*}=\lVert \tilde{T}^nHc-\tilde{T}^nHc^* \rVert\\
    &= \lVert d(\tilde T^n Hc, \tilde T^n Hc^*)\rVert
    \le \norm{B^n d(Hc, Hc^*)}
    \le \norm{B^n} \norm{d(Hc,Hc^*)} \\
    &\le \sigma^n \norm{d(Hc,Hc^*)}
\end{align*}
for large enough $n$. Therefore, $\rho(T^n c, c^*)\to 0$ as $n\to \infty$ and
\begin{equation*}
    \limsup_{n\to\infty}(\rho(T^nc,c^*))^{1/n}\le \sigma.
\end{equation*}
Sending $\sigma\downarrow r(B)$ yields the claim.

\medskip
\noindent\ref{item:existence3} To simplify notation, write $c^*=c$ and let $s(w,z)\coloneqq w-c(w,z)$. By \eqref{eq:bd_uprime}, we have $u'(c(\hat Y,\hat Z))\le u'(\hat Y)+M$, so Assumption \ref{asmp:Euler}\ref{item:Euler1} implies
\begin{equation*}
    A_z\coloneqq \E_z\left[\hat\beta\hat R\left\{u'(c(\hat Y,\hat Z))+v'(\hat Y)\right\}\right]<\infty.
\end{equation*}
Let $\bar A\coloneqq \max_{z\in \ZZ}A_z<\infty$. Since $u'(0)=\infty$, choose $\eta>0$ such that $u'(\eta)>\bar A$.

We first claim that
\begin{equation}
    s(w,z)=0 \qquad \text{whenever }w\le \eta. \label{eq:s=0}
\end{equation}
Suppose instead that $w\le \eta$ and $c(w,z)<w$. The borrowing constraint
is then slack, so the Euler equation holds with equality:
\begin{equation*}
    u'(c(w,z))=\E_z\left[\hat\beta\hat R\left\{u'(c(\hat w,\hat Z))+v'(\hat w)\right\}\right].
\end{equation*}
Since $c(\cdot,\hat z)$ is increasing and $u'$ and $v'$ are decreasing, we have
\begin{equation*}
    u'(c(\hat w,\hat Z))+v'(\hat w)\le u'(c(\hat Y,\hat Z))+v'(\hat Y).
\end{equation*}
Therefore,
\begin{equation*}
    u'(c(w,z))\le A_z\le \bar A<u'(\eta)\le u'(w).
\end{equation*}
On the other hand, since $c(w,z)<w$ and $u'$ is strictly decreasing, we have $u'(c(w,z))>u'(w)$, which is a contradiction. This proves \eqref{eq:s=0}.

We next claim that for $C\coloneqq u'(\eta)+M$, we have
\begin{equation}
    0 \le u'(c(w,z))s(w,z) \le Cw \qquad \text{for every }(w,z)\in \SS. \label{eq:u's_ub}
\end{equation}
Indeed, if $w\le \eta$, then $s(w,z)=0$ by \eqref{eq:s=0}. If $w>\eta$, then
\begin{equation*}
    u'(c(w,z)) \le u'(w)+M \le  u'(\eta)+M=C
\end{equation*}
by \eqref{eq:bd_uprime}, and $s(w,z)\le w$.

Let $B_t\coloneqq \prod_{i=1}^t\beta_i$ and $
Q_t\coloneqq \prod_{i=1}^t\beta_iR_i$, with the convention that empty products equal one. By $r(K(0))<1$, $r(K(1))<1$, and Lemma~\ref{lem:Lphi}\ref{item:Lphi1}, we have
\begin{align}
    M_0&\coloneqq \sum_{t=0}^{\infty}\max_{z\in \ZZ}\E_z B_t<\infty, &
    M_1&\coloneqq \sum_{t=0}^{\infty}\max_{z\in \ZZ}\E_z Q_t<\infty. \label{eq:M01}
\end{align}
Also set
\begin{equation}
    M_2\coloneqq \max_{z\in \ZZ}\E_z[\hat\beta\hat Y]<\infty. \label{eq:M2}
\end{equation}

Define the maximal wealth path $\set{\tilde{w}_t}_{t=0}^\infty$ by $\tilde{w}_0=w$ and $\tilde w_{t+1}=R_{t+1}\tilde w_t+Y_{t+1}$. Since consumption is nonnegative, $w_t\le \tilde w_t$ almost surely. Iterating the maximal-wealth recursion gives
\begin{equation*}
    \tilde w_t=\left(\prod_{i=1}^tR_i\right)w+\sum_{j=1}^tY_j\prod_{i=j+1}^tR_i.
\end{equation*}
Using Tonelli's theorem and the Markov property, we obtain
\begin{align*}
    \sum_{t=0}^{\infty}\E_{w,z}[B_t\tilde w_t]&=w\sum_{t=0}^{\infty}\E_z Q_t+\sum_{j=1}^{\infty}\E_z\left[B_jY_j\sum_{n=0}^{\infty}\E_{Z_j}\left(\prod_{\ell=1}^n\beta_\ell R_\ell\right)\right] \\
    &\le 
wM_1
+
M_1\sum_{j=1}^{\infty}\E_z[B_jY_j].
\end{align*}
Moreover,
\begin{equation*}
    \E_z[B_jY_j]=\E_z\left[B_{j-1}\E_{Z_{j-1}}[\hat\beta\hat Y]\right]\le M_2 \E_z B_{j-1}.
\end{equation*}
Combining this inequality with \eqref{eq:M01} and \eqref{eq:M2} yields
\begin{equation}
    \sum_{t=0}^{\infty}\E_{w,z}[B_t\tilde w_t]\le wM_1+M_0M_1M_2<\infty. \label{eq:sum_EBY}
\end{equation}
Since the terms in \eqref{eq:sum_EBY} are nonnegative, we have $\E_{w,z}[B_t\tilde w_t]\to 0$ as $t\to\infty$. Consequently, \eqref{eq:u's_ub} and $w_t\le \tilde w_t$ imply
\begin{equation*}
    0\le \E_{w,z}[B_tu'(c_t)(w_t-c_t)]\le C \E_{w,z}[B_tw_t]\le C \E_{w,z}[B_t\tilde w_t]\to 0,
\end{equation*}
so the transversality condition \eqref{eq:tvc} holds. By Proposition~\ref{prop:optimal}, $c$ is optimal.
\end{proof}

\subsection{Proof of Proposition \ref{prop:monotonew}}

We begin by defining
\begin{equation}
    \cC_0 = \set{c \in \cC \colon w \mapsto w - c(w,z) \text{ is increasing for all } z \in \ZZ}. \label{eq:cC0}
\end{equation}

\begin{lem}\label{lem:cC0}
$\cC_0$ is a closed subset of $\cC$, and $Tc \in \cC_0$ for all $c \in \cC_0$. 
\end{lem}

\begin{proof}
To see that $\cC_0$ is closed, let $\set{c_n}$ be a sequence in $\cC_0$, and suppose $c \in \cC$ satisfies $\rho(c_n, c) \to 0$. We claim that $c \in \cC_0$. To see this, for each $n$, the map $w \mapsto w - c_n(w,z)$ is increasing for all $z$, and the convergence $\rho(c_n,c) \to 0$ implies pointwise convergence $c_n (w,z) \to c(w,z)$ for all $(w,z) \in \SS$. Thus the monotonicity property is preserved in the limit, so $c \in \cC_0$.
	
Fix $c \in \cC_0$. We now show that $\xi \coloneqq Tc \in \cC_0$. By Lemma~\ref{lem:self_map}, $\xi \in \cC$, hence it suffices to show that $w \mapsto w - \xi(w,z)$ is increasing. Suppose not. Then there exist $z \in \ZZ$ and $w_1, w_2 \in (0, \infty)$ such that $w_1 < w_2$ and $w_1 - \xi(w_1, z) > w_2 - \xi (w_2, z)$. Since $w_1 - \xi (w_1, z) \ge 0$, $w_2 - \xi (w_2, z) \ge 0$ and $\xi(w_1,z) \le \xi (w_2, z)$ by Lemma~\ref{lem:self_map}, we must have $\xi(w_1, z) < w_1$ and  $\xi(w_1, z) < \xi(w_2, z)$. However, the definition of $T$ combined with the concavity of $u$ and $v$ then implies that
\begin{align*}
    (u' \circ \xi) (w_1, z) 
    &= \E_z \hat{\beta} \hat{R} (u'\circ c + v') (\hat{R}[w_1 - \xi(w_1,z)] + \hat{Y}, \hat{Z} ) \\
    &\le \E_z \hat{\beta} \hat{R} (u'\circ c + v') (\hat{R}[w_2 - \xi(w_2,z)] + \hat{Y}, \hat{Z} ) \le (u' \circ \xi) (w_2, z),
\end{align*}
which gives $\xi(w_1, z) \ge \xi(w_2, z)$, yielding a contradiction. Hence, $w \mapsto w - \xi (w,z)$ is increasing and $T$ is a self-map on $\cC_0$.
\end{proof}

\begin{proof}[Proof of Proposition~\ref{prop:monotonew}]
Since $T$ maps elements of the closed subset $\cC_0$ into itself by Lemma~\ref{lem:cC0}, Theorem~\ref{thm:existence} implies that $c^* \in \cC_0$. Hence, the stated claims hold.
\end{proof}

\subsection{Proof of Proposition \ref{prop:monotone_Yv}}

Let $T_j$ be the time iteration operator corresponding to specification $j\in\set{1,2}$. It suffices to show $T_1c \le T_2c$ for all $c \in \cC$. To see this, by Theorem~\ref{thm:existence}\ref{item:existence1}, we have $T_jc_1 \le T_jc_2$ whenever $c_1 \le c_2$. Thus, if $T_1c \le T_2c$ for all $c \in \cC$, then for any $c_1,c_2\in \cC$ with $c_1\le c_2$, we have $T_1c_1 \le T_1c_2 \le T_2c_2$. Iterating from any $c\in \cC$ and using Theorem~\ref{thm:existence}, we obtain
\begin{equation*}
    c_1^* = \lim_{n \to \infty}(T_1)^nc \le \lim_{n \to \infty}(T_2)^nc = c_2^*,
\end{equation*}
which proves the claim once $T_1c \le T_2c$ is established.
	
To show that $T_1c \le T_2c$ for any $c\in \cC$, take any $(w,z)\in \SS$ and define $\xi_j=(T_jc)(w,z)$. To show $\xi_1 \le \xi_2$, suppose to the contrary that $\xi_1 > \xi_2$. Since $Y_2\ge Y_1\ge m$, we have
\begin{equation*}
    \hat{R}(w-\xi_2)+\hat{Y}_2\ge \hat{R}(w-\xi_1)+\hat{Y}_1\ge m.
\end{equation*}
Because $c$ is increasing, $u',v_1',v_2'$ are decreasing, and $v_1'\ge v_2'$ on $[m,\infty)\cap (0,\infty)$, the definition of the time iteration operator in \eqref{eq:T_opr}--\eqref{eq:dom_T_opr} implies
\begin{align*}
    u'(\xi_2)>u'(\xi_1)&=\max \set{\E_z \hat{\beta} \hat{R}(u'\circ c + v_1') (\hat{R}(w - \xi_1) + \hat{Y}_1, \hat{Z}), u'(w)} \\
    &\ge \max \set{ \E_z \hat{\beta} \hat{R} (u'\circ c + v_2') (\hat{R}(w - \xi_2) + \hat{Y}_2, \hat{Z}), u'(w)} = u'(\xi_2),
\end{align*}
which is a contradiction. This completes the proof. \hfill \qedsymbol

\subsection{Proof of Proposition~\ref{prop:binding}}

Recall that, for all $c \in \cC$, $\xi(w,z) \coloneqq Tc(w,z)$ solves
\begin{equation}
    (u' \circ \xi)(w,z) = \max \set{ \E_z \hat{\beta} \hat{R} (u' \circ c + v') (\hat{R} [w - \xi(w,z)] + \hat{Y}, \hat{Z}), u'(w) }. \label{eq:T_opr_general}
\end{equation}
For each $z \in \ZZ$ and $c \in \cC$, define
\begin{equation*}
    \bar{w}_c (z) \coloneqq (u')^{-1}\left(\max\set{\E_z \hat{\beta} \hat{R} (u' \circ c + v') (\hat{Y}, \hat{Z}),u'(\infty)}\right)
\end{equation*}
and $\bar{w}(z) \coloneqq \bar{w}_{c^*} (z)$, which are well-defined in $(0,\infty]$ because $u'(0)=\infty$. If $\bar{w}_c(z)=\infty$, by definition we have
\begin{equation*}
    \E_z \hat{\beta} \hat{R} (u' \circ c + v') (\hat{Y}, \hat{Z})\le u'(\infty)<u'(w),
\end{equation*}
so \eqref{eq:T_opr_general} implies $\xi(w,z)=w$. Suppose $\bar{w}_c(z)<\infty$. We claim that $\xi(w,z)=w$ if and only if $w\le \bar{w}_c(z)$. If $\xi(w,z)<w$, then $(u' \circ \xi)(w,z) > u'(w)$. By \eqref{eq:T_opr_general}, we have
\begin{align*}
    u'(w) &< \E_z \hat{\beta} \hat{R} (u' \circ c + v') (\hat{R} [w - \xi(w,z) ] + \hat{Y}, \hat{Z}) \\
    &\le \E_z \hat{\beta} \hat{R} (u' \circ c + v') (\hat{Y}, \hat{Z}) = u'(\bar{w}_c (z)),
\end{align*}
so $w > \bar{w}_c (z)$. Taking the contrapositive, $w\le \bar{w}_c(z)$ implies $\xi(w,z)=w$. Conversely, if $\xi(w,z) = w$, then $(u' \circ \xi)(w,z) = u'(w)$. By \eqref{eq:T_opr_general}, we have
\begin{equation*}
    u'(w) \ge \E_z \hat{\beta} \hat{R} (u' \circ c + v')(\hat{Y}, \hat{Z}) = u'(\bar{w}_c (z)),
\end{equation*}
so $w \le \bar{w}_c (z)$. Setting $c=c^*$ and noting $\xi(w,z)=c^*(w,z)$ yields the claim. \hfill \qedsymbol

\section{Proof of Section~\ref{sec:asymptotic} results}

Unless otherwise stated, Assumptions~\ref{asmp:U}--\ref{asmp:power_u} are maintained. In the proof of Theorem~\ref{thm:MPC1}, Assumption~\ref{asmp:power_u} is replaced by the power utility specification stated in that theorem, which permits $\psi=0$. The proof has three ingredients. Lemma~\ref{lem:Tc_bd} shows how asymptotic MPC bounds propagate under time iteration. Lemma~\ref{lem:mpc_ub_n} iterates the upper bound from the maximal policy $c(w,z)=w$. Lemma~\ref{lem:upc_ub} identifies the limiting scaled marginal utility when $\delta<\gamma$. These results are then combined with lower bounds for the case $\delta\ge \gamma$.

\subsection{Common asymptotic lemmas}

\begin{lem}[Asymptotic MPC bounds propagation under time iteration]\label{lem:Tc_bd}
Let $c \in \cC$ and $F$ be as in \eqref{eq:F}.
\begin{enumerate}
    \item\label{item:Tc_ub} If $\delta<\gamma$, assume $\Pr_z(\hat{\beta}\hat{R}>0)>0$ for all $z\in \ZZ$. If $x\in [1,\infty]^\ZZ$ and
    \begin{equation*}
		\limsup_{w \to \infty} \frac{c(w,z)}{w} \le x(z)^{-1/\gamma} 
	\end{equation*}
    for all $z\in \ZZ$, then
	\begin{equation}
		\limsup_{w \to \infty} \frac{Tc(w,z)}{w} \le (Fx)(z)^{-1/\gamma}  \qquad \text{for all $z\in \ZZ$.} \label{eq:Tc_ub}
	\end{equation}
    \item\label{item:Tc_lb} Suppose $x\in [1,\infty)^\ZZ$, $\delta\ge \gamma$, and there exists $m>0$ such that $R_t \in \set{0} \cup [m,\infty)$ almost surely. If
    \begin{equation*}
		\liminf_{w \to \infty} \frac{c(w,z)}{w} \ge x(z)^{-1/\gamma} 
	\end{equation*}
    for all $z\in \ZZ$, then
	\begin{equation}
		\liminf_{w \to \infty} \frac{Tc(w,z)}{w} \ge (Fx)(z)^{-1/\gamma}  \qquad \text{for all $z\in \ZZ$.} \label{eq:Tc_lb}
	\end{equation}
\end{enumerate}
Consequently, under the assumptions of case \ref{item:Tc_lb}, if
\begin{equation*}
	\lim_{w \to \infty} \frac{c(w,z)}{w}=x(z)^{-1/\gamma} 
\end{equation*}
for all $z\in \ZZ$, then
\begin{equation}
	\lim_{w \to \infty} \frac{Tc(w,z)}{w} = (Fx)(z)^{-1/\gamma}  \qquad \text{for all $z\in \ZZ$.} \label{eq:Tc_lim}
	\end{equation}
\end{lem}

\begin{proof}
\ref{item:Tc_ub}
Fix $z\in \ZZ$ and let $\alpha = \limsup_{w \to \infty}Tc(w,z)/w$. By definition, we can take a sequence $\set{w_n}$ with $w_n\to\infty$ such that $\alpha = \lim_{n \to \infty} Tc(w_n,z)/w_n$. Define
\begin{equation}
	\alpha_n = \frac{Tc(w_n, z)}{w_n} \in (0,1]
	\quad \text{and} \quad 
	\lambda_n = \frac{c(\hat{R} (1-\alpha_n) w_n + \hat{Y}, \hat{Z})}{w_n} >0. \label{eq:lambdan}
\end{equation}
Let us show that 
\begin{equation}\label{eq:ls_ub}
	\limsup_{n \to \infty} \lambda_n \le x(\hat{Z})^{-1/\gamma} \hat{R} (1-\alpha).
\end{equation}
To see this, if $\alpha < 1$ and $\hat{R} > 0$, then since $\hat{R} (1-\alpha_n) w_n \to \hat{R} (1-\alpha) \cdot \infty = \infty$, by assumption we have
\begin{align*}
	\limsup_{n \to \infty}\lambda_n&= \limsup_{n \to \infty} \frac{c(\hat{R} (1-\alpha_n) w_n + \hat{Y},\hat{Z})}{\hat{R} (1-\alpha_n) w_n + \hat{Y}}\left(\hat{R}(1-\alpha_n) + \frac{\hat{Y}}{w_n}\right)  \\
	&\le \limsup_{w \to \infty} \frac{c(w,\hat{Z})}{w} \hat{R}(1-\alpha)\le x(\hat{Z})^{-1/\gamma} \hat{R} (1-\alpha),
\end{align*}
which is \eqref{eq:ls_ub}. If $\alpha = 1$ or $\hat{R} = 0$, then 
\begin{align*}
	\lambda_n &= \frac{c(\hat{R} (1-\alpha_n) w_n + \hat{Y},\hat{Z})}{\hat{R} (1-\alpha_n) w_n + \hat{Y}}\left(\hat{R}(1-\alpha_n) + \frac{\hat{Y}}{w_n}\right)  \\
	&\le \hat{R}(1-\alpha_n) + \frac{\hat{Y}}{w_n} \to \hat{R} (1-\alpha) =0,
\end{align*}
so again \eqref{eq:ls_ub} holds. 
	
Since $\xi_n \coloneqq Tc(w_n,z) = \alpha_n w_n$ solves the Euler equation \eqref{eq:foc}, by Assumption~\ref{asmp:power_u} and the definition of $\lambda_n$, we have
\begin{align}
	0 &= \frac{u'(\alpha_n w_n)}{u'(w_n)} - \max \set{\E_z \hat{\beta} \hat{R} \frac{(u'\circ c + v')(\hat{R} (1-\alpha_n)w_n + \hat{Y}, \hat{Z})}{u'(w_n)}, 1} \notag \\
	&= \alpha_n^{-\gamma} - \max \set{\E_z \hat{\beta} \hat{R} \left(\lambda_n^{-\gamma} + \psi \frac{(\hat{R} (1-\alpha_n)w_n + \hat{Y})^{-\delta}}{w_n^{-\gamma}} \right), 1}. \label{eq:alphan}
\end{align}
Therefore,
\begin{equation*}
	\alpha_n^{-\gamma} \ge \E_z \hat{\beta} \hat{R} \left(\lambda_n^{-\gamma} + \psi \frac{(\hat{R} (1-\alpha_n)w_n + \hat{Y})^{-\delta}}{w_n^{-\gamma}}	\right).
\end{equation*}
Letting $n \to \infty$, applying Fatou's lemma, and using the convention in Footnote \ref{fn:K}, we obtain
\begin{align}
	\alpha^{-\gamma} &= \lim_{n \to \infty} \alpha_n^{-\gamma} \ge \liminf_{n \to \infty} \E_z \hat{\beta} \hat{R} \left(\lambda_n^{-\gamma} + \psi \frac{(\hat{R} (1-\alpha_n)w_n + \hat{Y})^{-\delta}}{w_n^{-\gamma}}\right) \notag \\
	&\ge \E_z \hat{\beta} \hat{R} \left((\limsup_{n \to \infty} \lambda_n)^{-\gamma} + \lim_{n \to \infty} \psi \frac{(\hat{R} (1-\alpha_n)w_n + \hat{Y})^{-\delta}}{w_n^{-\gamma}}\right) \notag \\
	&\ge \E_z \hat{\beta} \hat{R} \left(\left[x(\hat{Z})^{-1/\gamma} \hat{R}(1-\alpha)\right]^{-\gamma} + \lim_{n \to \infty} \psi (\hat{R} (1-\alpha_n)w_n + \hat{Y})^{-\delta}w_n^\gamma\right) \notag \\
    &=[K(1-\gamma)x](z)(1-\alpha)^{-\gamma}+L_v, \label{eq:Fatou}
\end{align}
where
\begin{equation}
    L_v\coloneqq \psi \E_z\left[\hat{\beta}\hat{R}\lim_{n \to \infty} (\hat{R} (1-\alpha_n) + \hat{Y}/w_n)^{-\delta}w_n^{\gamma-\delta}\right]\ge 0. \label{eq:Lv}
\end{equation}
\begin{itemize}
    \item If $\delta>\gamma$, \eqref{eq:Fatou} implies
    \begin{equation*}
        \alpha^{-\gamma}\ge [K(1-\gamma)x](z)(1-\alpha)^{-\gamma}.
    \end{equation*}
    Solving the inequality for $\alpha$ yields
    \begin{equation*}
        \limsup_{w\to\infty}\frac{Tc(w,z)}{w}=\alpha\le \left(1+[K(1-\gamma)x](z)^{1/\gamma}\right)^{-1},
    \end{equation*}
    which is equivalent to \eqref{eq:Tc_ub}.
    \item If $\delta=\gamma$, the limit inside \eqref{eq:Lv} equals
    \begin{equation*}
        \lim_{n \to \infty} (\hat{R} (1-\alpha_n) + \hat{Y}/w_n)^{-\gamma}=
        [\hat{R}(1-\alpha)]^{-\gamma},
    \end{equation*}
    so \eqref{eq:Fatou} implies
    \begin{equation*}
        \alpha^{-\gamma}\ge [K(1-\gamma)(x+\psi 1)](z)(1-\alpha)^{-\gamma}.
    \end{equation*}
    Solving the inequality for $\alpha$ yields
    \begin{equation*}
        \limsup_{w\to\infty}\frac{Tc(w,z)}{w}=\alpha\le \left(1+[K(1-\gamma)(x+\psi 1)](z)^{1/\gamma}\right)^{-1},
    \end{equation*}
    which is equivalent to \eqref{eq:Tc_ub}.
    \item If $\delta<\gamma$, the limit inside \eqref{eq:Lv} equals
    \begin{equation*}
        \lim_{n \to \infty} (\hat{R} (1-\alpha_n) + \hat{Y}/w_n)^{-\delta}w_n^{\gamma-\delta}=[\hat{R}(1-\alpha)]^{-\delta}\cdot\infty=\infty.
    \end{equation*}
    Since $\Pr_z(\hat{\beta}\hat{R}>0)>0$, \eqref{eq:Fatou} implies $\alpha^{-\gamma}\ge \infty$ and hence $\alpha=0$, which is equivalent to \eqref{eq:Tc_ub}.
\end{itemize}

\medskip
\noindent \ref{item:Tc_lb} 
If $\Pr_z(\hat{\beta}\hat{R} > 0) = 0$, then $Tc(w,z) \equiv w$ and the result is immediate. In what follows, we assume $\Pr_z(\hat{\beta}\hat{R} > 0) > 0$. 

Fix $z\in \ZZ$ and let $\alpha = \liminf_{w \to \infty}Tc(w,z)/w$. Since $c(w,z)\le w$ and $\Pr_z(\hat{\beta}\hat{R} > 0) > 0$, part \ref{item:Tc_ub} implies
\begin{equation*}
    \limsup_{w\to\infty}\frac{Tc(w,z)}{w}\le (F1)(z)^{-1/\gamma}<1,
\end{equation*}
so $\alpha<1$. Choose a sequence $\set{w_n}$ with $w_n\to\infty$ such that $Tc(w_n,z)/w_n\to \alpha$, and let $\alpha_n,\lambda_n$ be as in case \ref{item:Tc_ub}. Using the same argument as in the derivation of \eqref{eq:ls_ub}, we obtain
\begin{equation}
	\liminf_{n \to \infty} \lambda_n \ge x(\hat{Z})^{-1/\gamma} \hat{R} (1-\alpha) \label{eq:ls_lim}
\end{equation}
and \eqref{eq:alphan} continues to hold. Because the integrand vanishes when $\hat{R}=0$, it suffices to consider $\hat{R}>0$. Then, by assumption, $\hat{R}\ge m>0$, so \eqref{eq:ls_lim} implies
\begin{equation*}
	\liminf_{n\to\infty} \lambda_n\ge x(\hat{Z})^{-1/\gamma}m(1-\alpha).
\end{equation*}
Since $x(z)>0$ for all $z$ and $\ZZ$ is a finite set, for any
\begin{equation*}
	\ubar{\lambda}\in \left(0,\min_{z\in \ZZ}x(z)^{-1/\gamma}m(1-\alpha)\right)
\end{equation*}
and $\bar{\alpha}\in (\alpha,1)$, there exists $N\in \N$ such that $\lambda_n\ge \ubar{\lambda}$, $\alpha_n<\bar{\alpha}$, and $w_n\ge 1$ for all $n\ge N$ and $\hat{Z}\in \ZZ$. Using $\delta\ge \gamma$, the integrand in the expectation in \eqref{eq:alphan} is bounded above by
\begin{equation*}
	\hat{\beta}\hat{R}\left(\ubar{\lambda}^{-\gamma}+\psi (m(1-\bar{\alpha}))^{-\delta}\right),
\end{equation*}
which is integrable. Since $\alpha_n<\bar{\alpha}<1$, the Euler equation is interior. Letting $n \to \infty$ in \eqref{eq:alphan}, applying the reverse Fatou lemma, and using $\alpha<1$, we obtain
\begin{align*}
	\alpha^{-\gamma} &=\lim_{n \to \infty} \E_z \hat{\beta} \hat{R} \left(\lambda_n^{-\gamma} + \psi \frac{(\hat{R} (1-\alpha_n)w_n + \hat{Y})^{-\delta}}{w_n^{-\gamma}}\right)\\
	&\le \E_z \hat{\beta} \hat{R} \left(\left[x(\hat{Z})^{-1/\gamma} \hat{R}(1-\alpha)\right]^{-\gamma} + \lim_{n \to \infty} \psi \frac{(\hat{R} (1-\alpha_n)w_n + \hat{Y})^{-\delta}}{w_n^{-\gamma}}\right).
\end{align*}
The rest of the proof is the same as case \ref{item:Tc_ub}.
\end{proof}

\begin{lem}[Iterated upper bound for the asymptotic MPC]\label{lem:mpc_ub_n}
If $\delta<\gamma$, assume $\Pr_z(\hat{\beta}\hat{R}>0)>0$ for all $z\in \ZZ$. Define the sequence $\set{x_n}\subset [1,\infty]^\ZZ$ by $x_0=1$ and $x_n=F^nx_0$, where $F$ is as in \eqref{eq:F}. Then
\begin{equation}
    \limsup_{w\to\infty}\frac{c^*(w,z)}{w}\le x_n(z)^{-1/\gamma} \qquad \text{for all $z\in \ZZ$.} \label{eq:mpc_ub_n}
\end{equation}
\end{lem}

\begin{proof}
Define the sequence $\set{c_n} \subset \cC$ by $c_0(w,z) = w$ and $c_n = Tc_{n-1}$ for all $n \ge 1$. Since $Tc(w,z) \le w$ for any $c\in \cC$, we have $c_1(w,z) = Tc_0(w,z) \le w = c_0(w,z)$. Since $T:\cC \to \cC$ is monotone, by induction $c_n \le c_{n-1}$ for all $n$ and $c^*(w,z) = \lim_{n \to \infty} c_n(w,z)$, where $c^*$ is the unique fixed point of $T$ by Theorem~\ref{thm:existence}.
	
By the definitions of $c_0$ and $x_0$, we have $c_0(w,z)/w = 1 = x_0(z)^{-1/\gamma}$, so
\begin{equation*}
	\limsup_{w \to \infty} \frac{c_0(w,z)}{w} \le x_0(z)^{-1/\gamma}
\end{equation*}
for all $z\in\ZZ$. Since $c_n \downarrow c^* \ge 0$ pointwise, a repeated application of Lemma~\ref{lem:Tc_bd} gives 
\begin{equation*}
	\limsup_{w\to \infty} \frac{c^*(w,z)}{w} \le \limsup_{w\to \infty} \frac{c_n(w,z)}{w} \le x_n(z)^{-1/\gamma},
\end{equation*}
which is \eqref{eq:mpc_ub_n}.
\end{proof}

\subsection{Proof of Theorem \ref{thm:MPC0}}

\begin{lem}[Convergence of scaled marginal utility]
\label{lem:upc_ub}
Suppose $\delta<\gamma$ and $\Pr_z(\hat\beta\hat R>0)>0$ for all $z\in\ZZ$. If $K(1-\gamma)$ and $K(1-\delta)$ are irreducible with $r(K(1-\gamma)),r(K(1-\delta))<1$, and $Y_t\ge m$ almost surely for some $m>0$, then the function
\begin{equation}\label{eq:f_func}
    f(w,z) \coloneqq
    w^\delta (u'\circ c^*)(w,z)
\end{equation}
is bounded on $[m,\infty)\times\ZZ$. Moreover, for all $z\in\ZZ$,
\begin{equation*}
    \lim_{w\to\infty}f(w,z)=v^*(z)\in(0,\infty),
\end{equation*}
where $v^*\coloneqq [I-K(1-\delta)]^{-1}K(1-\delta)(\psi1)$.
\end{lem}

\begin{proof}
Since $\delta<\gamma$, by \eqref{eq:F} we have $F1=\infty$ entrywise. Hence, Lemma~\ref{lem:mpc_ub_n}, applied with $n=1$, gives
\begin{equation}\label{eq:mpc0_for_f}
    \lim_{w\to\infty}\frac{c^*(w,z)}{w}=0
    \qquad \text{for all $z\in\ZZ$.}
\end{equation}
We first establish boundedness of $f$. Choose $\epsilon\in(0,1)$ sufficiently
small that
\begin{equation*}
    M_\gamma\coloneqq(1-\epsilon)^{-\gamma}K(1-\gamma) \quad\text{and}\quad M_\delta\coloneqq(1-\epsilon)^{-\delta}K(1-\delta)
\end{equation*}
satisfy $r(M_\gamma),r(M_\delta)<1$. By the Perron-Frobenius theorem, there exist $v_\gamma,v_\delta\in \R_{++}^\ZZ$ such that
\begin{equation*}
    M_\gamma v_\gamma=r(M_\gamma)v_\gamma \quad\text{and}\quad M_\delta v_\delta=r(M_\delta)v_\delta.
\end{equation*}
Moreover, by Lemma~\ref{lem:G_oper}, the operator
$G_\epsilon:\R_+^\ZZ\to\R_+^\ZZ$ defined by
$G_\epsilon v\coloneqq M_\delta(v+\psi1)$ has the unique fixed point
\begin{equation*}
    v_\epsilon\coloneqq (I-M_\delta)^{-1}M_\delta(\psi1).
\end{equation*}
For $C>0$, define $g_C:[m,\infty)\to\R_+^\ZZ$ by
\begin{equation*}
    g_C(w)\coloneqq Cv_\gamma w^{-\gamma}+(v_\epsilon+v_\delta)w^{-\delta}.
\end{equation*}
Because $Y_t\ge m$ almost surely, $\SS_0\coloneqq[m,\infty)\times\ZZ$ is
an invariant state space. We therefore apply the preceding fixed-point arguments
on $\SS_0$: because $\SS_0$ is invariant, the restriction of $c^*$ to $\SS_0$ is a fixed point of the restricted time iteration
operator. The proof of Theorem~\ref{thm:existence} applies verbatim on $\SS_0$, so this fixed point is unique. In particular, if
\begin{equation*}
    C\ge C_0\coloneqq\max_{z\in\ZZ}\frac{1}{v_\gamma(z)},
\end{equation*}
then $g_C\in\hH$ on this restricted state space.

Let
\begin{equation*}
    q_C(w,z)\coloneqq(u')^{-1}(g_{C,z}(w))
    \quad\text{and}\quad
    c_C(w,z)\coloneqq(u')^{-1}[(\tilde T g_C)_z(w)].
\end{equation*}
By Lemma~\ref{lem:conjug}, $c_C=Tq_C$. Since $g_C(w,z)\ge w^{-\gamma}$,
we have $q_C(w,z)\le w$. Hence Lemma~\ref{lem:Tc_bd}\ref{item:Tc_ub},
applied with $x=1$, implies
\begin{equation*}
    0\le \limsup_{w\to\infty}\frac{c_{C_0}(w,z)}{w}\le (F1)(z)^{-1/\gamma}=0,
\end{equation*}
so the limit is zero for all $z\in \ZZ$. Since $\ZZ$ is finite, there exists $W\ge m$ such that
\begin{equation}\label{eq:cC_large_w}
    c_{C_0}(w,z)\le\epsilon w
    \qquad \text{for all $w\ge W$ and $z\in\ZZ$.}
\end{equation}
If $C\ge C_0$, then $g_C\ge g_{C_0}$. The monotonicity of $\tilde T$ and
the fact that $(u')^{-1}$ is decreasing imply $c_C\le c_{C_0}$, so
\eqref{eq:cC_large_w} holds with $c_C$ in place of $c_{C_0}$.

It remains to obtain the same bound on $[m,W]$. For $z\in\ZZ$ and
$w\in[m,W]$, define
\begin{equation*}
    a_z(w)\coloneqq
    \E_z\hat\beta\hat Rv_\gamma(\hat Z)(\hat Rw+\hat Y)^{-\gamma}.
\end{equation*}
The function $a_z$ is continuous on $[m,W]$ by dominated convergence, since
\begin{equation*}
    0\le
    \hat\beta\hat Rv_\gamma(\hat Z)(\hat Rw+\hat Y)^{-\gamma}
    \le
    \left(\max_{\hat z\in\ZZ}v_\gamma(\hat z)\right)
    \hat\beta\hat R\hat Y^{-\gamma},
\end{equation*}
and the right-hand side is integrable by Assumption~\ref{asmp:Euler}\ref{item:Euler1}.
Moreover, $a_z(w)>0$ because $v_\gamma\gg0$ and
$\Pr_z(\hat\beta\hat R>0)>0$. Hence
\begin{equation*}
    a_0\coloneqq
    \min_{z\in\ZZ}\min_{w\in[m,W]}a_z(w)>0.
\end{equation*}
Putting $\hat w_C\coloneqq\hat R[w-c_C(w,z)]+\hat Y$, we have $\hat w_C\le\hat Rw+\hat Y$, and therefore
\begin{equation*}
    (\tilde T g_C)_z(w) \ge \E_z\hat\beta\hat R Cv_\gamma(\hat Z)\hat w_C^{-\gamma} \ge C a_z(w)\ge Ca_0.
\end{equation*}
Consequently,
\begin{equation*}
    c_C(w,z)\le(Ca_0)^{-1/\gamma}
    \qquad \text{for all $(w,z)\in[m,W]\times\ZZ$.}
\end{equation*}
Increasing $C\ge C_0$ if necessary, we may ensure that
$(Ca_0)^{-1/\gamma}\le\epsilon m$. Combining this inequality with
\eqref{eq:cC_large_w}, we obtain
\begin{equation}\label{eq:cC_global}
    c_C(w,z)\le\epsilon w
    \qquad \text{for all $(w,z)\in[m,\infty)\times\ZZ$.}
\end{equation}
We now verify that $\tilde T g_C\le g_C$. By the definition of $\tilde T$,
\begin{equation*}
    (\tilde T g_C)_z(w)=\max\set{A_{g_C}(w,z),w^{-\gamma}},
\end{equation*}
where
\begin{equation}
    A_{g_C}(w,z)
    \coloneqq\E_z\hat\beta\hat R
    \left[g_{C,\hat Z}(\hat w_C)+\psi\hat w_C^{-\delta}\right]. \label{eq:AgC}
\end{equation}
Since $C\ge C_0$, we have $w^{-\gamma}\le g_{C,z}(w)$. Moreover,
\eqref{eq:cC_global} implies
$\hat w_C\ge\hat R(1-\epsilon)w$. Since
\begin{equation*}
    \hat w_C^{-\gamma}
    \le \hat R^{-\gamma}(1-\epsilon)^{-\gamma}w^{-\gamma}, \qquad
    \hat w_C^{-\delta}
    \le \hat R^{-\delta}(1-\epsilon)^{-\delta}w^{-\delta},    
\end{equation*}
and integrands in \eqref{eq:AgC} vanish when $\hat R=0$, it follows that
\begin{align*}
    A_{g_C}(w,z)
    &\le Cw^{-\gamma}(M_\gamma v_\gamma)(z)+w^{-\delta}
    [M_\delta(v_\epsilon+v_\delta+\psi1)](z) \\
    &=Cw^{-\gamma}r(M_\gamma)v_\gamma(z)+w^{-\delta}[v_\epsilon+r(M_\delta)v_\delta](z) \\
    &\le Cv_\gamma(z)w^{-\gamma}+(v_\epsilon(z)+v_\delta(z))w^{-\delta}
    =g_{C,z}(w).
\end{align*}
Thus $\tilde T g_C\le g_C$. By the monotonicity of $\tilde T$,
Lemma~\ref{lem:conjug}, and Theorem~\ref{thm:existence},
\begin{equation*}
    (u'\circ c^*)(w,z)
    =\lim_{n\to\infty}(\tilde T^ng_C)_z(w)
    \le g_{C,z}(w).
\end{equation*}
It follows that
\begin{align}\label{eq:f_bounded}
    0<f(w,z)
    &\le Cv_\gamma(z)w^{\delta-\gamma}+v_\epsilon(z)+v_\delta(z) \notag \\
    &\le \max_{z\in\ZZ} \left(Cv_\gamma(z)m^{\delta-\gamma}+v_\epsilon(z)+v_\delta(z)\right)
    <\infty.
\end{align}
Hence $f$ is bounded on $[m,\infty)\times\ZZ$.

We next establish convergence. Define
\begin{equation*}
    \bar x(z)\coloneqq\limsup_{w\to\infty}f(w,z),
    \qquad
    \ubar x(z)\coloneqq\liminf_{w\to\infty}f(w,z),
\end{equation*}
and define $G:\R_+^\ZZ\to\R_+^\ZZ$ by
\begin{equation*}
    Gv\coloneqq K(1-\delta)(v+\psi1).
\end{equation*}
By Lemma~\ref{lem:G_oper}, $G$ has the unique fixed point
\begin{equation*}
    v^*\coloneqq [I-K(1-\delta)]^{-1}K(1-\delta)(\psi1),
\end{equation*}
and $G^nv\to v^*$ for every $v\in\R_+^\ZZ$.

By \eqref{eq:mpc0_for_f}, the borrowing constraint is slack for all sufficiently
large $w$. The Euler equation and the power utility specification then give
\begin{equation}\label{eq:f_recursion}
    f(w,z)=\E_z\hat\beta\hat R
    (\hat{w}/w)^{-\delta}
    [f(\hat w,\hat Z)+\psi],
    \qquad
    \hat w=\hat R[w-c^*(w,z)]+\hat Y.
\end{equation}
Fix $z\in\ZZ$, choose an increasing sequence $w_n\to\infty$ such that
$f(w_n,z)\to\ubar x(z)$, and let $\hat w_n\coloneqq\hat R[w_n-c^*(w_n,z)]+\hat Y$. If $\hat R>0$, \eqref{eq:mpc0_for_f} implies
$\hat w_n/w_n\to\hat R$ and $\hat w_n\to\infty$. If $\hat R=0$, the integrand in \eqref{eq:f_recursion} is zero. Therefore,
Fatou's lemma yields
\begin{align*}
    \ubar x(z)
    &\ge \E_z\hat\beta\hat R
    \liminf_{n\to\infty}
    (\hat{w}_n/w_n)^{-\delta}
    [f(\hat w_n,\hat Z)+\psi] \\
    &\ge \E_z\hat\beta\hat R^{1-\delta}
    [\ubar x(\hat Z)+\psi]
    =(G\ubar x)(z).
\end{align*}
Hence $\ubar x\ge G\ubar x$.

Next choose an increasing sequence $w_n\to\infty$ such that
$f(w_n,z)\to\bar x(z)$. Let $M<\infty$ be the uniform bound in
\eqref{eq:f_bounded}. Fix $\eta\in(0,1)$. By \eqref{eq:mpc0_for_f}, for all
sufficiently large $n$,
$c^*(w_n,z)\le\eta w_n$. Since $\hat w_n\ge\hat Y\ge m$, we have
$f(\hat w_n,\hat Z)\le M$, and
\begin{equation*}
    \frac{\hat w_n}{w_n}\ge(1-\eta)\hat R.
\end{equation*}
Consequently, the integrand in \eqref{eq:f_recursion} is bounded above by
\begin{equation*}
    (1-\eta)^{-\delta}(M+\psi)
    \hat\beta\hat R^{1-\delta},
\end{equation*}
with the integrand interpreted as zero if $\hat R=0$ by the convention in Footnote \ref{fn:K}. This upper bound is
integrable because $K(1-\delta)$ has finite entries. The reverse Fatou lemma
therefore gives
\begin{align*}
    \bar x(z)
    &\le \E_z\hat\beta\hat R
    \limsup_{n\to\infty}
    (\hat{w}_n/w_n)^{-\delta}
    [f(\hat w_n,\hat Z)+\psi] \\
    &\le \E_z\hat\beta\hat R^{1-\delta}
    [\bar x(\hat Z)+\psi]
    =(G\bar x)(z).
\end{align*}
Hence $\bar x\le G\bar x$.

Iterating these two inequalities and using Lemma~\ref{lem:G_oper}, we obtain
\begin{equation*}
    \bar x\le G^n\bar x\to v^*,
    \qquad
    \ubar x\ge G^n\ubar x\to v^*.
\end{equation*}
Thus $\bar x\le v^*\le\ubar x$. Since $\ubar x\le\bar x$ by definition,
$\ubar x=\bar x=v^*$. Therefore,
\begin{equation*}
    \lim_{w\to\infty}f(w,z)=v^*(z)
    \qquad \text{for all $z\in\ZZ$.}
\end{equation*}
Finally, $v^*\gg0$ because $\psi>0$ and
$\Pr_z(\hat\beta\hat R>0)>0$ for all $z$, completing the proof.
\end{proof}

\begin{proof}[Proof of Theorem~\ref{thm:MPC0}]
Let $\set{x_n}$ be as in Lemma \ref{lem:mpc_ub_n}. Since $\delta<\gamma$, the
definition of $F$ in \eqref{eq:F} gives $x_n(z)=\infty$ for all $n\ge 1$
and $z\in\ZZ$. Hence, Lemma~\ref{lem:mpc_ub_n} implies
\begin{equation*}
	\lim_{w \to \infty} \frac{c^*(w,z)}{w} = 0.
\end{equation*}
In particular, $c^*(w,z)<w$ for large enough $w$, so the Euler equation \eqref{eq:foc} implies
\begin{equation*}
	(u'\circ c^*)(w,z)=\E_z\hat{\beta}\hat{R}(u'\circ c^*+v')(\hat{w},\hat{Z})\ge \E_z\hat{\beta}\hat{R}v'(\hat{w}).
\end{equation*}
Using the power utility form and noting that $\hat{w}=\hat{R}[w-c^*(w,z)]+\hat{Y}\le \hat{R}w+\hat{Y}$, we obtain
\begin{equation*}
	c^*(w,z)^{-\gamma}\ge \E_z\hat{\beta}\hat{R}\psi(\hat{R}w+\hat{Y})^{-\delta} \iff c^*(w,z)\le \left(\psi\E_z\hat{\beta}\hat{R}(\hat{R}w+\hat{Y})^{-\delta}\right)^{-1/\gamma}.
\end{equation*}
Dividing both sides by $w^{\delta/\gamma}$ gives
\begin{equation*}
    \frac{c^*(w,z)}{w^{\delta/\gamma}}\le\left(\psi\E_z\hat\beta\hat R(\hat R+\hat{Y}/w)^{-\delta}\right)^{-1/\gamma}.
\end{equation*}
For $w\ge 1$, we have $(\hat R+\hat{Y}/w)^{-\delta} \ge (\hat R+\hat Y)^{-\delta}$. Moreover,
\begin{equation*}
    0<d(z)\coloneqq \psi\E_z\hat\beta\hat R(\hat R+\hat Y)^{-\delta}\le \psi\E_z\hat\beta\hat R\hat Y^{-\delta}=\E_z\hat\beta\hat R v'(\hat Y)<\infty,
\end{equation*}
where strict positivity follows from
$\Pr_z(\hat\beta\hat R>0)>0$ and finiteness follows from Assumption~\ref{asmp:Euler}\ref{item:Euler1}. Therefore,
\begin{equation*}
    \limsup_{w\to\infty}\frac{c^*(w,z)}{w^{\delta/\gamma}}\le d(z)^{-1/\gamma}<\infty
\end{equation*}
and \eqref{eq:MPC0_and_speed_conv} holds.

Under the additional assumptions in the second statement of the theorem, noting that $u'(c)=c^{-\gamma}$, Lemma~\ref{lem:upc_ub} implies
\begin{equation*}
    \lim_{w\to\infty}\frac{c^*(w,z)}{w^{\delta/\gamma}}
    =\lim_{w\to\infty}(w^\delta(u'\circ c^*)(w,z))^{-1/\gamma}=v^*(z)^{-1/\gamma},
\end{equation*}
so  \eqref{eq:speed_conv} holds.
\end{proof}

\subsection{Proof of Proposition~\ref{prop:MPC0}}

Put $K \coloneqq K(1-\gamma)$. Suppose first that $\delta<\gamma$. Since $K$ is irreducible, every row of $K$ is nonzero. Therefore, $\Pr_z(\hat{\beta}\hat{R}>0)>0$ for all $z\in \ZZ$ and the claim follows from Theorem~\ref{thm:MPC0}. Suppose next that $\delta\ge \gamma$. Since $K$ is irreducible and $r(K)\ge 1$, Lemma \ref{lem:F_iter}\ref{item:F_iter2} (applied to $A=K(1-\gamma)$ and $b=K(1-\gamma)\psi 1$ or $b=0$ depending on $\delta=\gamma$ or $\delta>\gamma$) implies $F^n1\to\infty$ entrywise. Letting $n\to\infty$ in Lemma \ref{lem:mpc_ub_n}, we obtain
\begin{equation*}
    0\le \limsup_{w\to\infty}\frac{c^*(w,z)}{w}\le (F^n1)^{-1/\gamma}\to 0,
\end{equation*}
implying $c^*(w,z)/w\to 0$. \hfill \qedsymbol

\subsection{Proof of Theorem \ref{thm:MPC1}}

Let $x^*=x_1^*$ be the unique fixed point of $F$ in $\R_+^\ZZ$ established in Lemma \ref{lem:F_iter}, which clearly satisfies $x^*\ge 1$. Let $\bar{c}(z) = x^*(z)^{-1/\gamma} \in (0,1]$ and consider the subset
\begin{equation*}
	\cC_1 = \set{ c\in \cC : \text{$c(w,z) \ge \bar{c}(z)w$ for all $w>0$ and $z\in \ZZ$}}.
\end{equation*}
Let us show that $T\cC_1 \subset \cC_1$. By definition, we need to show $\xi\coloneqq Tc(w,z)\ge \bar{c}(z)w$ for all $c\in \cC_1$, $w>0$, and $z\in \ZZ$.

Suppose first that $\bar{c}(z)=1$. Then $Fx^*=x^*$ implies
\begin{equation*}
    [K(1-\gamma)(x^*+\psi 1)](z)=0.
\end{equation*}
Since $x^*\gg 0$, the $z$-th row of $K(1-\gamma)$ must be zero, implying $\Pr_z(\hat{\beta}\hat{R}>0)=0$. Consequently, $Tc(w,z)=w=\bar{c}(z)w$ for all $w>0$ and the claim holds.

Suppose next that $\bar{c}(z)<1$. If $\xi\ge \bar{c}(z)w$, there is nothing to prove, so assume $\xi < \bar{c}(z) w$. Then the Euler equation is interior. Since $c(\hat{w},\hat{Z})\ge \bar{c}(\hat{Z})\hat{w}$, $\hat{w}\ge \hat{R}(1-\bar{c}(z))w$, and $\delta=\gamma$, we have
\begin{align*}
    \xi^{-\gamma}&=\E_z\hat{\beta}\hat{R}\left(c(\hat{w},\hat{Z})^{-\gamma}+\psi \hat{w}^{-\gamma}\right)
    \le \E_z\hat{\beta}\hat{R}\left(\bar{c}(\hat{Z})^{-\gamma}+\psi\right)\hat{w}^{-\gamma}\\
    &\le \E_z\hat{\beta}\hat{R}^{1-\gamma}\left(\bar{c}(\hat{Z})^{-\gamma}+\psi\right)(1-\bar{c}(z))^{-\gamma}w^{-\gamma}\\
    &=[K(1-\gamma)(x^*+\psi1)](z)(1-\bar{c}(z))^{-\gamma}w^{-\gamma}
    =x^*(z)w^{-\gamma}=[\bar{c}(z)w]^{-\gamma},
\end{align*}
where the penultimate equality uses $Fx^*=x^*$. Therefore, $\xi\ge \bar{c}(z)w$ and the claim holds.

Let $\set{c_n}\subset \cC$ be as in the proof of Lemma \ref{lem:mpc_ub_n}. Letting $n \to \infty$ in \eqref{eq:mpc_ub_n} yields
\begin{equation}\label{eq:limsup_bd}
	\limsup_{w \to \infty} \frac{c^*(w,z)}{w} \le x^*(z)^{-1/\gamma}=\bar{c}(z)
\end{equation}
for all $z\in \ZZ$. Since $c_0(w,z)\coloneqq w\in \cC_1$ and $T\cC_1\subset \cC_1$, we have $\set{c_n}\subset \cC_1$. Since $c_n \to c^*$ pointwise, passing to the limit gives
\begin{equation*}
	\liminf_{w\to \infty} \frac{c^*(w,z)}{w} \ge \bar{c}(z)=x^*(z)^{-1/\gamma}
\end{equation*}
for all $z\in \ZZ$. Combining this inequality with \eqref{eq:limsup_bd} yields \eqref{eq:MPC1}, as required.  \hfill \qedsymbol

\subsection{Proof of Theorem \ref{thm:MPC2}}

Consider the utility function of wealth $v_1(w)\coloneqq \psi m_2^{\gamma-\delta}p(w;\gamma)$, and let $c_1^*$ be the corresponding optimal consumption function. Since $\delta\ge \gamma$, for $w\ge m_2$, we have
\begin{equation*}
    v'(w)=\psi w^{-\delta}\le \psi m_2^{\gamma-\delta}w^{-\gamma}=v_1'(w).
\end{equation*}
Since $Y\ge m_2$, we have $\E_z\hat{\beta}\hat{R}v_1'(Y)\le \E_z\hat{\beta}\hat{R}\psi m_2^{-\delta}<\infty$ and Assumption \ref{asmp:Euler} holds. Proposition~\ref{prop:monotone_Yv} then implies $c^*(w,z)\ge c_1^*(w,z)$ for all $w\ge m_2$ and $z$. Consider the following operators:
\begin{enumerate}
	\item\label{item:F1} Let $F_1$ be the $F$ in \eqref{eq:F} and $T_1$ be the time iteration operator $T$ for the parameters $(\gamma,\delta,\psi)=(\gamma,\gamma,\psi m_2^{\gamma-\delta})$. Moreover, let $x_1^*$ be the unique fixed point of $F_1$.
	\item\label{item:F2} Let $F_2$ be the $F$ in \eqref{eq:F} and $T_2$ be the time iteration operator $T$ for the parameters $(\gamma,\delta,\psi)$. Let $x_2^*$ be the unique fixed point of $F_2$. Define the sequence $\set{x_n}\subset \R_+^{\ZZ}$ by $x_0=x_1^*$ and $x_n=F_2x_{n-1}$.
\end{enumerate}
By \ref{item:F2}, $c^*$ is the unique fixed point of $T_2$ in $\cC$. Using the monotonicity of $T$, we obtain from applying $T_2^n$ to both sides of $c^*\ge c_1^*$ that
\begin{equation}\label{eq:linf}
	\liminf_{w\to\infty}\frac{c^*(w,z)}{w} = \liminf_{w\to\infty}\frac{T_2^n c^*(w,z)}{w} \ge \liminf_{w\to\infty}\frac{T_2^n c_1^*(w,z)}{w}.
\end{equation}
Moreover, by \ref{item:F1}, $c_1^*$ is the unique fixed point of $T_1$ in $\cC$. By Theorem~\ref{thm:MPC1}, we have
\begin{equation*}
	\lim_{w\to \infty} \frac{c_1^*(w,z)}{w} = x_1^*(z)^{-1/\gamma} = x_0(z)^{-1/\gamma}.
\end{equation*}
Applying \eqref{eq:Tc_lim} in Lemma \ref{lem:Tc_bd} iteratively, we obtain 
\begin{equation*}
	\lim_{w\to\infty}\frac{T_2^n c_1^*(w,z)}{w} = x_n(z)^{-1/\gamma}.
\end{equation*}
Combined with \eqref{eq:linf}, this implies that
\begin{equation*}
	\liminf_{w\to\infty}\frac{c^*(w,z)}{w} \ge x_n(z)^{-1/\gamma}
\end{equation*}
for all $n \ge 0$. Letting $n\to\infty$ and using convergence of $\set{x_n}$ to $x_2^*$, we obtain
\begin{equation*}
	\liminf_{w\to\infty}\frac{c^*(w,z)}{w}\ge x_2^*(z)^{-1/\gamma}.
\end{equation*}
Furthermore, by Lemmas \ref{lem:mpc_ub_n} and \ref{lem:F_iter}, we have
\begin{equation*}
    \limsup_{w\to\infty}\frac{c^*(w,z)}{w}\le \lim_{n\to\infty} (F_2^n1)(z)^{-1/\gamma}=x_2^*(z)^{-1/\gamma},
\end{equation*}
which completes the proof of \eqref{eq:MPC2}. \hfill \qedsymbol

\printbibliography
	
\end{document}